\input harvmac.tex

\let\includefigures=\iftrue
\let\useblackboard=\iftrue
\newfam\black

%Figure Stuff
\includefigures
\message{If you do not have epsf.tex (to include figures),}
\message{change the option at the top of the tex file.}
\input epsf
\def\figin{\epsfcheck\figin}\def\figins{\epsfcheck\figins}
\def\epsfcheck{\ifx\epsfbox\UnDeFiNeD
\message{(NO epsf.tex, FIGURES WILL BE IGNORED)}
\gdef\figin##1{\vskip2in}\gdef\figins##1{\hskip.5in}% blank space instead
\else\message{(FIGURES WILL BE INCLUDED)}%
\gdef\figin##1{##1}\gdef\figins##1{##1}\fi}
\def\DefWarn#1{}
\def\figinsert{\goodbreak\midinsert}
\def\ifig#1#2#3{\DefWarn#1\xdef#1{fig.~\the\figno}
\writedef{#1\leftbracket fig.\noexpand~\the\figno}%
\figinsert\figin{\centerline{#3}}\medskip\centerline{\vbox{
\baselineskip12pt\advance\hsize by -1truein
\noindent\footnotefont{\bf Fig.~\the\figno:} #2}}
\bigskip\endinsert\global\advance\figno by1}
%%%
\else
\def\ifig#1#2#3{\xdef#1{fig.~\the\figno}
\writedef{#1\leftbracket fig.\noexpand~\the\figno}%
%\figinsert\figin{\centerline{#3}}\medskip
%\centerline{\vbox{\baselineskip12pt
%\advance\hsize by -1truein\noindent
%\footnotefont{\bf Fig.~\the\figno:} #2}}
%\bigskip\endinsert
\global\advance\figno by1}
\fi

\def\doublefig#1#2#3#4{\DefWarn#1\xdef#1{fig.~\the\figno}
\writedef{#1\leftbracket fig.\noexpand~\the\figno}%
\figinsert\figin{\centerline{#3\hskip1.0cm#4}}\medskip\centerline{\vbox{
\baselineskip12pt\advance\hsize by -1truein
\noindent\footnotefont{\bf Fig.~\the\figno:} #2}}
\bigskip\endinsert\global\advance\figno by1}

\def\mathboxit#1{\vbox{\hrule\hbox{\vrule\kern8pt\vbox{\kern8pt
\hbox{$\displaystyle #1$}\kern8pt}\kern8pt\vrule}\hrule}}

%%BLACKBOARD FONT STUFF
\useblackboard
\message{If you do not have msbm (blackboard bold) fonts,}
\message{change the option at the top of the tex file.}
\font\blackboard=msbm10 scaled \magstep1
\font\blackboards=msbm7
\font\blackboardss=msbm5
\textfont\black=\blackboard
\scriptfont\black=\blackboards
\scriptscriptfont\black=\blackboardss
\def\Bbb#1{{\fam\black\relax#1}}
\else
\def\Bbb{\bf}
\fi
% *************************************
%\draft
%

\def\yboxit#1#2{\vbox{\hrule height #1 \hbox{\vrule width #1
\vbox{#2}\vrule width #1 }\hrule height #1 }}
\def\fillbox#1{\hbox to #1{\vbox to #1{\vfil}\hfil}}
\def\ybox{{\lower 1.3pt \yboxit{0.4pt}{\fillbox{8pt}}\hskip-0.2pt}}
%
%
%%MATH MACROS
%Greek letters and their bars
\def\ep{\epsilon}

%More bars

%\def\l{\left}
%\def\r{\right}
\def\comments#1{}

\def\QC{\Bbb{C}}
\def\QD{\Bbb{D}}
\def\QH{\Bbb{H}}

\def\QR{\Bbb{R}}
\def\QS{\Bbb{S}}

\def\QZ{\Bbb{Z}}

\def\Re{{\rm Re\hskip0.1em}}
\def\Im{{\rm Im\hskip0.1em}}

\def\CC{{\cal C}}

\def\CF{{\cal F}}

\def\CO{{\cal O}}%AEL

\def\CV{{\cal V}}

%AEL

\def\ta{\tilde\alpha}

\def\II{\relax{I\kern-.10em I}}

\def\IZ{\relax\ifmmode\mathchoice
{\hbox{\cmss Z\kern-.4em Z}}{\hbox{\cmss Z\kern-.4em Z}}
{\lower.9pt\hbox{\cmsss Z\kern-.4em Z}}
{\lower1.2pt\hbox{\cmsss Z\kern-.4em Z}}
\else{\cmss Z\kern-.4emZ}\fi}
\def\IB{\relax{\rm I\kern-.18em B}}
\def\IC{{\relax\hbox{$\inbar\kern-.3em{\rm C}$}}}
\def\ID{\relax{\rm I\kern-.18em D}}
\def\IE{\relax{\rm I\kern-.18em E}}
\def\IF{\relax{\rm I\kern-.18em F}}
\def\IG{\relax\hbox{$\inbar\kern-.3em{\rm G}$}}
\def\IGa{\relax\hbox{${\rm I}\kern-.18em\Gamma$}}
\def\IH{\relax{\rm I\kern-.18em H}}
\def\II{\relax{\rm I\kern-.18em I}}
\def\IK{\relax{\rm I\kern-.18em K}}
\def\IP{\relax{\rm I\kern-.18em P}}
%\def\IX{\relax{\rm X\kern-.01em X}}
%this doesn't work

%

\def\inbar{\,\vrule height1.5ex width.4pt depth0pt}

\font\cmss=cmss10 
\def\IR{\relax{\rm I\kern-.18em R}}

\def\BR{\IR}
\def\BZ{Z} % for now

\def\BR{\IR}

\def\lp10{\ell_p^{10}}
\def\lp11{\ell_p^{11}}
\def\R11{R_{11}}

\def\frac#1#2{{#1 \over #2}}

%%simeon's macros
%\def\pp{\partial}
%\def\uu{^}
%\def\ll{_}
%\def\a{\alpha}
%\def\b{\beta}
%\def\s{\sigma}
%\def\g{\gamma}
%\def\st{^*}

%\def\e{\epsilon}
%\def\th{\theta}
%\def\thb{\bar{\theta}}
%\def\psb{\bar{\psi}}
%\def\adot{{\dot{\alpha}}}
%\def\bdot{{\dot{\beta}}}

%\def\x{\chi}
%\def\m{\mu}
%\def\n{\nu}
%\def\d{\delta}

%\def\p{\pi}

%\def\r{\rho}
%\def\t{\tau}
%\def\sq{\xi}

%\def\k{\kappa}

%\def\G{\Gamma}

%\def\L{\Lambda}
%%\def\S{\Sigma} %%this is a bad idea
%%No it's not.  And stop commenting out
%%my macros without permission, john
%% -SH
%% like i said before it IS a bad idea to mess with harvmac.

\def\1dag{^{1\dagger}}
\def\2dag{^{2\dagger}}

\def\R#1#2#3{{{R_{#1}}^{#2}}_{#3}}

\def\ot{\otimes}

%%note the switch!!

  %%%% some words are hard to type.
\def\Ga{\Gamma}
%%ENGLISH MACROS

\def\ie{{\it i.e.}}
\def\cf{{\it c.f.}}

\hyphenation{Di-men-sion-al}

%%REFERENCING MACROS

%%

\def\ra{\rightarrow}
\def\pa{\partial}
\def\bz{{\bar z}}
\def\al{\alpha}

%\McGreevyKB
\lref\McGreevyKB{
J.~McGreevy and H.~Verlinde,
``Strings from Tachyons,''
[arXiv:hep-th/0304224].
%%CITATION = HEP-TH 0304224;%%
}

%\MartinecKA
\lref\MartinecKA{
E.~J.~Martinec,
``The annular report on non-critical string theory,''
[arXiv:hep-th/0305148].
%%CITATION = HEP-TH 0305148;%%
}

%\KMS
\lref\KMS{
I.R.~Klebanov, J. Maldacena, N. Seiberg,
``D-brane Decay in Two-Dimensional String Theory''
[arXiv:hep-th/0305159].
}

%\CLQDbranes
\lref\CLQDbranes{
J.~McGreevy, J.~Teschner H.~Verlinde,
``Classical and Quantum D-branes in 2D String Theory,''
[arXiv:hep-th/0305194].
%%CITATION = HEP-TH 0304224;%%
}

\lref\AlexandrovNN{
S.~Y.~Alexandrov, V.~A.~Kazakov and D.~Kutasov,
``Non-perturbative effects in matrix models and D-branes,''
[arXiv:hep-th/0306177].
%%CITATION = HEP-TH 0306177;%%
}

\lref\TakayanagiSM{
T.~Takayanagi and N.~Toumbas,
``A matrix model dual of type 0B string theory in two dimensions,''
[arXiv:hep-th/0307083].
%%CITATION = HEP-TH 0307083;%%
}

\lref\DouglasUP{
M.~R.~Douglas, I.~R.~Klebanov, D.~Kutasov, J.~Maldacena, 
E.~Martinec and N.~Seiberg,
``A New Hat For The c=1 Matrix Model,''
[arXiv:hep-th/0307195].
%%CITATION = HEP-TH 0307195;%%
}

\lref\GaiottoYF{
D.~Gaiotto, N.~Itzhaki and L.~Rastelli,
``On the BCFT Description of Holes in the c=1 Matrix Model,''
[arXiv:hep-th/0307221].
%%CITATION = HEP-TH 0307221;%%
}

\lref\emiltwo{
E.~Martinec, 
``Defects, decay, and dissipated states,'' 
[arXiv:hep-th/0210231].}

\lref\DO{
H.~Dorn and H.~J.~Otto,
``Two and three point functions in Liouville theory,''
Nucl.\ Phys.\ B {\bf 429}, 375 (1994)
[arXiv:hep-th/9403141].
%%CITATION = HEP-TH 9403141;%%
}

\lref\FZZ{
V.~Fateev, A.~B.~Zamolodchikov and A.~B.~Zamolodchikov,
``Boundary Liouville field theory. I: Boundary state and boundary  two-point function,''
[arXiv:hep-th/0001012].
%%CITATION = HEP-TH 0001012;%%
}

\lref\PST{
B.~Ponsot, V.~Schomerus and J.~Teschner,
``Branes in the Euclidean AdS(3),''
JHEP {\bf 0202}, 016 (2002)
[arXiv:hep-th/0112198].
%%CITATION = HEP-TH 0112198;%%
}

\lref\Hosomichi{
K. Hosomichi,
``Bulk-Boundary Propagator in Liouville Theory on a Disc,''
JHEP {\bf 0111}, 044 (2001)
[arXiv:hep-th/0108093]
}

\lref\PonTe{
B.~Ponsot and J.~Teschner,
``Boundary Liouville field theory: Boundary three point function,''
Nucl.\ Phys.\ B {\bf 622}, 309 (2002)
[arXiv:hep-th/0110244].
%%CITATION = HEP-TH 0110244;%%
}

\lref\qPonTe{
B.~Ponsot and J.~Teschner,
"Clebsch-Gordan and Racah-Wigner coefficients for a continuous
series of representations of $U_q(sl(2,R))$",
Comm.\ Math. Phys. 224 (2001) 613-655
[arXiv:math.QA/0007097]
}

\lref\qhyp{
M. Nishizawa, K. Ueno,
"Integral solutions of q-hypergeometric difference systems
with $|q|=1$",
[arXiv:q-alg/9612014]
}

\lref\TLbound{
J.~Teschner,
``Remarks on Liouville theory with boundary,''
[arXiv:hep-th/0009138].
%%CITATION = HEP-TH 0009138;%%
}

\lref\TLtwo{
J.~Teschner,
``A lecture on the Liouville vertex operators,''
[arXiv:hep-th/0303150].
%%CITATION = HEP-TH 0303150;%%
}

\lref\TB{
J.~Teschner,
``Liouville theory revisited,''
Class.\ Quant.\ Grav.\  {\bf 18}, R153 (2001)
[arXiv:hep-th/0104158].
%%CITATION = HEP-TH 0104158;%%
}

\lref\SeiLiou{
N. Seiberg,
``Notes on quantum Liouville theory and quantum gravity'',
Progr. Theor. Phys. Suppl. {\bf 102} (1990) 319
}

\lref\ZZ{
A.~B.~Zamolodchikov and A.~B.~Zamolodchikov,
``Structure constants and conformal bootstrap in Liouville field theory,''
Nucl.\ Phys.\ B {\bf 477}, 577 (1996)
[arXiv:hep-th/9506136].
%%CITATION = HEP-TH 9506136;%%
}

\lref\ZZB{
A.~B.~Zamolodchikov and A.~B.~Zamolodchikov,
``Liouville field theory on a pseudosphere,''
[arXiv:hep-th/0101152].
%%CITATION = HEP-TH 0101152;%%
}

%\GinspargIS
\lref\GinspargIS{
P.~Ginsparg and G.~W.~Moore,
``Lectures On 2-D Gravity And 2-D String Theory,''
[arXiv:hep-th/9304011].
%%CITATION = HEP-TH 9304011;%%
}

\lref\igor{
I.~R.~Klebanov,
``String theory in two-dimensions,''
[arXiv:hep-th/9108019].
%%CITATION = HEP-TH 9108019;%%
}

\lref\Kondo{ 
I.~Affleck, A.W.W.~ Ludwig,
``Exact conformal field theory results on the multichannel Kondo effect:
Single-fermion Green's functions, self-energy, and resistivity,'' 
Phys. Rev. {\bf B48} (1993) 7297
}

\lref\boundpert{
A.~Recknagel, D.~Roggenkamp, V.~Schomerus,
``On relevant boundary perturbations of unitary minimal models''
Nucl.\ Phys.\ {\bf B588} (2000) 552-564
[arXiv:hep-th/0003110].
}

\lref\cardylec{
J. Cardy,
``Conformal invariance and statistical mechanics'',
Lectures at Les Houches 1988
}

\lref\Barnes{
E.W.~Barnes, 
``Theory of the double gamma function'', 
Phil.\ Trans.\ Roy.\ Soc.\ {\bf A196} (1901) 265--388
}

\lref\Shintani{
T.~Shintani, 
``On a Kronecker limit formula for real quadratic fields,'' 
J.\ Fac.\ Sci.\ Univ.\
Tokyo Sect.\ 1A Math.\ {\bf 24} (1977) 167--199
}

\lref\Senone{
A. Sen,
``Rolling Tachyon'', JHEP {\bf 04} (2002) 048 
[arXiv:hep-th/0203211]
}

\lref\Sentwo{
A. Sen,
``Time evolution in open string theory'', JHEP {\bf 10} (2002) 003 
[arXiv:hep-th/0207105]
}

\lref\SeiShen{
N. Seiberg, S. Shenker,
``Note on background (in)dependence'', Phys. Rev. {\bf D45} (1992) 4581
}

\lref\GutperleXF{
M.~Gutperle and A.~Strominger,
``Timelike boundary Liouville theory,''
Phys. Rev. {\bf D67} (2003) 126002,
[arXiv:hep-th/0301038].
%%CITATION = HEP-TH 0301038;%%
}

\lref\LLM{
N.~Lambert, H.~Liu and J.~Maldacena, ``Closed strings from
decaying D-branes,'' [arXiv:hep-th/0303139].
}
%%CITATION = HEP-TH 0303139;%%

\Title{\vbox{\baselineskip3pt
\hbox{SFB 288 Preprint}}}
{\vbox{ \centerline{On Boundary Perturbations in Liouville Theory}}}
\vskip -19pt
{\vbox{ \centerline{\titlefont and}}}
\vskip 15pt
{\vbox{ \centerline{\titlefont Brane Dynamics in Noncritical String Theories}}}

%{\vbox{ \centerline{\titlefont in Noncritical String Theories}}}

\bigskip

\bigskip
\bigskip
\centerline{J\"org Teschner}
\bigskip
\bigskip
\bigskip
\centerline{{\it Institut f\"ur theoretische Physik, Freie Universit\"at
Berlin, 14195 Berlin}}
\bigskip
\bigskip
\centerline{\bf Abstract}
\bigskip
\noindent
We study certain relevant boundary perturbations of Liouville theory
and discuss implications of our results for the brane dynamics in
noncritical string theories. Our results include

(i) There exist monodromies in the parameter $\mu_{\rm B}$ 
of the Neumann-type
boundary condition that can
create an admixture represented by the
Dirichlet type boundary condition, for example.

(ii) Certain renormalization group flows can be 
studied perturbatively,
which allows us to determine the results of 
the corresponding brane decays.

(iii) There exists a simple renormalization group flow
that can be calculated exactly. In all the cases that
we have studied the renormalization group flow
acts like a covering transformation for the monodromies 
mentioned under (i).

\bigskip
\Date{August, 2003}

\newsec{Introduction}
\def\mub{\mu_{\rm B}}
\def\murb{\mu_{\rm B,ren}}
\def\De{\Delta}
\def\be{\beta}
\def\de{\delta}
\def\ga{\gamma}
\def\si{\sigma}
\def\la{\lambda}

In the present paper we start investigating perturbations
of quantum Liouville theory by operators that break conformal
invariance on the boundary of the domain on which 
Liouville theory is defined. 

\subsec{Motivation}
Our motivation is three-fold:

\item{\bf 1.} 
There has recently been important progress 
concerning the role of D-branes in the dualities between
noncritical string theories and matrix models, exhibiting 
these dualities as holographic open/closed string
dualities 
\refs{\McGreevyKB,\MartinecKA,
\KMS,\CLQDbranes,\AlexandrovNN,\GaiottoYF,
      \TakayanagiSM,\DouglasUP}. 
However, the picture that is emerging from 
\McGreevyKB-\DouglasUP\
does not encompass the D-branes defined by Neumann-type
boundary conditions for the Liouville direction so far. The 
implications of the fact 
that one of the most basic observables of the matrix model, the macroscopic
loop operator, is related to D-instantons 
with Neumann type boundary conditions 
for the Liouville direction \FZZ\ do not seem to 
be fully understood from the point of view of D-brane physics in 
noncritical string theories yet.
Important clues in this direction can be expected to come from 
an improved understanding of the D-branes in 
noncritical string theories, their dynamics and their mutual
relations. One of our results appears
to be quite intriguing in this respect: 
Recall that the Neumann type boundary conditions 
in Liouville theory are
parametrized by the so-called boundary cosmological constant 
$\mu_{\rm B}$. Analytic continuation
w.r.t. $\mub$ can create an admixture
represented by the Dirichlet type boundary condition\foot{This was
previously observed in \MartinecKA\ on the level of 
the boundary states. Here we show that one indeed finds complete
decoupling also between the open string sectors associated to the
two components.}.

\item{\bf 2.} The present understanding of time-dependent processes
in string theory like the decay of D-branes does not seem
to be satisfactory yet.
Currently there are two main approaches for studying decay processes
triggered by tachyonic modes on D-branes. In the first of these
approaches one forgets about the time-direction and studies the
renormalization group (RG) flows induced by perturbing the
resulting boundary CFT with relevant operators. It is generally 
believed that the fixed points of the 
boundary RG flow represent possible end-points of the decay of an
unstable brane, see e.g. \emiltwo\ for a discussion. 
In a second approach to the description of brane decay processes
one tries to construct an exact time dependent solution of the
open string theory, \ie\ a boundary CFT that involves the CFT for 
the time-like $X^0$-coordinate non-trivially \refs{\Senone,\Sentwo}.
We would like to point out that the 
noncritical string theories may serve as a useful toy model for 
exploring certain aspects of time-dependent processes
in string theory. One has a free parameter which determines the 
speed of decay of the brane. Processes which take place slowly can 
here be studied perturbatively, unlike the examples studied in 
\refs{\Senone,\Sentwo}.

\item{\bf 3.} 
From a somewhat more general perspective it seems
important to gain some insight into the consequences of breaking
conformal invariance on the boundary in 
conformal field theories that have a continuous spectrum.
In the example that we study in this paper we will observe a
remarkable new phenomenon: The couplings that correspond to 
generic relevant boundary fields have a trivial
scale dependence only, they are ``frozen''. The exceptions include 
those boundary fields which create {\it normalizable} states
from the vacuum. Adapting the terminology of \SeiLiou\ to the 
present situation we will call these fields ``macroscopic''. 
Perturbing with macroscopic fields will generate nontrivial RG flows which
may have nontrivial infrared fixed points.

\subsec{Overview}

We begin our discussion in Section 2 by providing the necessary
results from boundary Liouville theory. 
Of particular importance for us will be to understand
(a) the analytic properties of the theory w.r.t. the
boundary parameter 
$\mu_{\rm B}$, and 
(b) the operator product expansion of boundary fields.
The dependence of the theory w.r.t. $\mu_{\rm B}$ is found to be quite 
interesting: We will see that boundary Liouville theory lives on 
a certain branched cover of the complex $\mu_{\rm B}$-plane.
A special case of 
the corresponding covering transformations turns out 
to be related to the remarkable relation between Dirichlet 
and Neumann type boundary conditions that we had 
mentioned above.
We present and discuss the relevant results in Section 2, 
with derivations given in the appendices.

In Section 3 we discuss some examples for RG flows generated 
by perturbing Liouville theory with relevant boundary fields. 
In the case that we perturb with nearly marginal boundary fields
one may use boundary perturbation theory to calculate the
relevant beta-function and to determine the new fixed point
that the theory flows to. We 
also discuss the simple soluble example of the 
RG flow which removes the admixture of Dirichlet-type 
boundary condition which is produced by performing a monodromy 
w.r.t. the parameter $\mub$ of the Neumann type boundary conditions.
In all the cases that we were able to treat we 
observe that the renormalization group flow acts like 
a covering transformation which relates different Neumann-type
boundary conditions with the same value of $\mub$.

The results of Section 3 are interpreted from the perspective of
two-dimensional string theory in Section 4. The RG flows 
generated by the macroscopic boundary fields are interpreted 
in terms of brane decays that take place spontaneously, 
whereas the trivial scale dependence of the 
couplings for microscopic boundary fields
is related to brane decays that are triggered by strong 
external sources. The latter always lead to the complete 
disappearance
of the brane, whereas the former produces a 
different brane with the same value of $\mub$. 
The soluble RG flow mentioned in the previous paragraph 
has an explicit time-dependent description in terms of the 
boundary states introduced in \refs{\Senone,\Sentwo}.

The appendices  contain the derivations of some technical results
which are interesting in their own right. This includes in particular
a detailed study of the analytic properties of the structure 
functions, and an analysis of the operator product 
expansion of boundary fields in boundary Liouville theory.

\subsec{Conventions}
We will assume that the reader is familiar with Liouville theory 
in the case of periodic boundary conditions on the cylinder, as
discussed in \TB\ and references therein. 
Our conventions will follow those of 
\refs{\ZZ,\TB}, with the following exception.
Let $|\al\rangle$ be the state generated by acting with
the primary field $V_{\al}(z,\bz)$ on the vacuum, 
\eqn\ketaldef{
|\al\rangle\,:=\,\lim_{z\ra 0}V_{\al}(z,\bz)|0\rangle.
}
We then define $|P\rangle$ by 
\eqn\ketPdef{
|P\rangle\,=\,\frac{1}{\sqrt{2\pi}}\,|\al\rangle,\quad{\rm if}\;\;
\al=\frac{Q}{2}+iP,
}
which is normalized such that $\langle P'|P\rangle=\de(P'-P)$ if 
$P,P'\in\BR$.

\newsec{More on boundary Liouville theory}

As a preparation we will need to develop our understanding of quantum 
Liouville theory with conformally invariant boundary conditions 
\refs{\FZZ,\TLbound,\ZZB,\PonTe} 
a little further. To be specific, let us consider 
Liouville theory on the upper half plane, with 
certain boundary conditions imposed along the real axis.
For the reader's convenience let
us begin by reviewing the necessary 
results from \refs{\FZZ,\TLbound,\ZZB,\PonTe}.

\subsec{Boundary conditions}

Liouville theory permits two types of boundary conditions
which preserve conformal invariance: One being of Dirichlet type \ZZB, 
the other being a generalization of Neumann type boundary conditions 
\refs{\FZZ,\TLbound,\PonTe}.

The Dirichlet type boundary condition is classically defined
by the requirement that the classical Liouville field $\varphi(z,\bz)$
diverges near the real axis like $-2\log\Im(z)$. 
The corresponding boundary condition for quantum Liouville 
theory may be characterized by 
the boundary state $\langle B_{\rm D}|$, which we will write as
\eqn\ZZstate{ 
\langle B_{\rm D}\,| 
\;=\;\int_{0}^{\infty} \!dP\;\Psi_{\rm ZZ}(P)\,
\langle\!\langle P\,|\;,\qquad
\Psi_{\rm ZZ}(P)
\;=\;-2^{\frac{3}{4}}\,v(P)\, ,
}
where 
$|P\rangle\!\rangle$ is the 
Ishibashi-state built from $|P\rangle$,
and the function $v(P)$ is defined by
\eqn\vpdef{
v(P)\,:=\,
\bigl(\pi\mu\ga(b^2)\bigr)^{-\frac{iP}{b}}
\frac{-4\pi iP}{\Ga(1-2ibP)\Ga(1-2iP/b)}\;.
}

The second important class of conformally invariant boundary conditions for 
Liouville theory may be defined in the semi-classical limit
$c\ra\infty$ by imposing the boundary condition
\eqn\Nbc{
i\big(\,\pa-\bar{\pa}\,\big)\,\varphi\,=\,4\pi \mub b^2\,e^{\frac{\varphi}{2}}.
}
along the real axis.
%\eqn\yyy{
%S_{\rm bound}\;=\;\int_{\pa\Ga}\frac{d\tau}{2\pi} \;g^{\frac{1}{4}}\bigl(
%QK+2\pi \mu_{\rm B}e^{b\varphi}\bigr),
%}
The parameter $\mub$ which labels the boundary conditions
is called the boundary cosmological constant.
The corresponding boundary states were constructed in \FZZ.
They can be represented as
\eqn\boundst{
\langle B_\si| 
\;=\;\frac{1}{2\pi}\int_{\QS}d\al\;A^{\si}_{\al}\;
\langle\!\langle \al|\;,
}
where $\QS=\frac{Q}{2}+i\QR^+$, 
the parameter $\si$ used in \boundst\ is related to the
boundary cosmological constant $\mub$ via
\eqn\smub{
\cos \pi b(2\si-Q) 
=\frac{\mub}{\sqrt{\mu}}\sqrt{\sin\pi b^2}\;,
}
and the one-point function $A^{\si}_{\al}$ is given as
\eqn\onept{
A^{\si}_{\al}\;=\;
\sqrt{\frac{\pi}{2}}
\frac{\cosh(2\pi P(2\si-Q))}{\sinh 2\pi bP\sinh 2\pi b^{-1}P}
\Psi_{\rm ZZ}(P)\quad{\rm if}\;\,\al=\frac{Q}{2}+iP.
}

We will see later that correlation functions of boundary Liouville theory 
have nice analytic properties in their dependence on 
the boundary parameter $\si$. The relation \smub\ uniformizes 
the branched cover of the complex $\mub$-plane on which 
boundary Liouville theory can be defined.
To begin with, we will consider the case where
the parameter $\si$ takes values in $\QS$. We notice that we then have
a one-to-one correspondence between the boundary conditions
labelled by $\sigma\in \QS:=
\frac{Q}{2}+i\QR^+$ and the spectrum of Liouville theory with 
periodic boundary conditions (``Cardy-case'').

\subsec{Spectrum I - Continuous part}

Let us consider Liouville theory on the strip
$[0,\pi]$ with boundary conditions labelled by
parameters $\sigma_2$ and $\sigma_1$. The spectrum 
$\CH_{\sigma_2\sigma_1}^{\rm B}$
of Liouville theory
on the strip has been 
determined in \TLbound. It may be represented as 
\eqn\spec{
\CH_{\sigma_2\sigma_1}^{\rm B}\;=\;\int_{\QS} d\al\; \CV_{\al,c},
}
where $\QS=\frac{Q}{2}+i\BR^+$ and
$\CV_{\al,c}$ is the highest weight representation of the Virasoro algebra
with weight $\De_{\alpha}=\al(Q-\al)$ and central charge $c=1+6Q^2$.
We will denote by $v_{\al}^{\sigma_2\sigma_1}$ the 
representative of the highest weight state of $\CV_{\al,c}$ in 
$\CH_{\sigma_2\sigma_1}^{\rm B}$,
normalized by
\eqn\bdnorm{
\langle\, v_{\al_2}^{\sigma_2\sigma_1}\, ,\, v_{\al_1}^{\sigma_2\sigma_1}\,
\rangle_{\CH_{\sigma_2\sigma_1}^{\rm B}}^{}\;=\;
\de_{\QS}^{}(\al_2-\al_1).
}
It is sometimes useful to note \TLbound\ that in the 
weak coupling asymptotics $\phi\rightarrow -\infty$ 
one may describe the states $v_{\al}^{\sigma_2\sigma_1}$
by the wave-functions $\psi_{\al}^{\si_2\si_1}(q)$, 
$q:= \int_{0}^{\pi}dx
\varphi(x)$,
such that
\eqn\wvfasym{
\psi_{\al}^{\si_2\si_1}(q)
\,\sim \,\frac{1}{\sqrt{2\pi}}e^{-\frac{Q}{2}q}
\big(e^{\al q}+r_{\al}^{\si_2\si_1}
e^{(Q-\al) q}\big)\quad{\rm for}\;\; q\rightarrow -\infty\;,
}
where the reflection amplitude $r_{\al}^{\si_2,\si_1}$ 
has an explicit 
expression given by 
\eqn\refamp{
\eqalign{ {}  r_{\al}^{\si_2\si_1}\,=\,&
\bigl(\pi \mu \ga(b^2) b^{2-2b^2}\bigr)^{\frac{Q-2\alpha}{2b}}\cr
&\times \frac{\Ga_b(2\alpha-Q)}{\Ga_b(Q-2\alpha)}
\frac{S_b(2Q-\si_2-\si_1-\alpha)S_b(\si_2+\si_1-\al)}
     {S_b(\al+\si_1-\si_2)S_b(\al+\si_2-\si_1)}
\;.
}}
Definitions and relevant properties of the special 
functions $\Ga_b$ and
$S_b$ are reviewed in the Appendix A.1. For the moment
let us simply note that $\Ga_b(x)$ and
$S_b(x)$ are analytic in the strip $0<\Re(x)<Q$ and 
have a simple pole
at $x=0$. 

\subsec{Boundary fields}

The states $v_{P}^{\sigma_2\sigma_1}$ are 
created from the vacuum by the
boundary fields $\Psi^{\sigma_2\sigma_1}_{\al}(x)$, $\al=\frac{Q}{2}+iP$. 
These boundary fields are fully characterized by the three point function
\eqn\threept{
\eqalign{
\langle\, 
\Psi^{\sigma_1\sigma_3}_{\alpha_3}(x_3) 
\Psi^{\sigma_3\sigma_2}_{\alpha_2}(x_2)
\Psi^{\sigma_2\sigma_1}_{\alpha_1}(x_1)\rangle\,
=& \,|x_3-x_2|^{\De_{\al_1}-\De_{\al_3}-\De_{\al_2}}
|x_3-x_1|^{\De_{\al_2}-\De_{\al_3}-\De_{\al_1}}\times\cr 
\times & \;
|x_2-x_1|^{\De_{\al_3}-\De_{\al_2}-\De_{\al_1}}
C_{\alpha_{3}\alpha_{2}\alpha_{1}}^{\sigma_{3}\sigma_{2}\sigma_{1}}
}}
The  explicit formula for 
$C_{\alpha_{3}\alpha_{2}\alpha_{1}}^{\sigma_{3}\sigma_{2}\sigma_{1}}
\equiv
C_{Q-\alpha_{3}|\alpha_{2}\alpha_{1}}^{\sigma_{3}\sigma_{2}\sigma_{1}}$
found in \PonTe\ is reviewed and further studied in Appendix A. 

An explicit description for the operator product expansion of 
boundary operators can be found by combining the results of 
\PonTe\ and \TLtwo:
\eqn\OPE{
\eqalign{
\Psi^{\sigma_3\sigma_2}_{\alpha_2}(x_2)& \Psi^{\sigma_2\sigma_1}_{\alpha_1}(x_1)= \cr
&= \int_{\QS}d\alpha_3\;
C_{\alpha_{3}|\alpha_{2}\alpha_{1}}^{\sigma_{3}\sigma_{2}\sigma_{1}}
\,|x_2-x_1|^{\De_{\alpha_3}-\De_{\alpha_2}-\De_{\alpha_1}}\,
\Psi^{\sigma_3\sigma_1}_{\alpha_3}(x_1)\;+\;{\rm descendants},
}}
The expansion \OPE\ is applicable if $|\Re(\al_1-\al_2)|<Q/2$
and $|\Re(\al_1+\al_2-Q)|<Q/2$. The general case can be obtained
from \OPE\ by means of analytic continuation, see Appendix C
for more details.

For these and other purposes it turns out to be rather useful 
to change the normalization of the boundary fields via
\eqn\bdrenorm{
\Phi^{\sigma_2\sigma_1}_{\alpha}(x)\;=\;g_{\al}^{\si_2\si_1} 
\Psi^{\sigma_2\sigma_1}_{\alpha}(x),
}
where the function $g_{\al}^{\si_2\si_1}$ will be chosen as \PonTe\
\eqn\gfktdef{
\eqalign{
g_{\al}^{\si_2\si_1}= &
\frac{\big(\pi \mu\ga(b^2)b^{2-2b^2}\big)^{\frac{\al}{2b}}}
     {\Ga_b(2Q-\al-\si_1-\si_2)} \cr
& \times\frac{\Ga_b(Q)\Ga_b(Q-2\al)\Ga_b(2\si_1)\Ga_b(2Q-2\si_2)}
      {\Ga_b(Q-\al+\si_1-\si_2)\Ga_b(Q-\al+\si_2-\si_1)\Ga_b(\si_1+\si_2-\al)}.
}}
Using the fields $\Phi^{\sigma_2\sigma_1}_{\alpha}(x)$ instead of 
$\Psi^{\sigma_2\sigma_1}_{\alpha}(x)$ has considerable 
advantages, for example:

\item{\bf 1.} The fields $\Phi^{\sigma_2\sigma_1}_{\alpha}(x)$ have a 
particularly simple behavior under the reflection $\al\rightarrow Q-\al$,
namely
\eqn\phirefl{
\Phi^{\sigma_2\sigma_1}_{\alpha}(x)\;=\;\Phi^{\sigma_2\sigma_1}_{Q-\alpha}(x).
}
This implies in particular that $\Phi^{\sigma_2\sigma_1}_{\alpha}(x)$ 
is non-vanishing at $\al=Q/2$ (unlike $\Psi^{\sigma_2\sigma_1}_{\alpha}(x)$).
In the case $c=25 \Leftrightarrow b=1$ this implies furthermore
that the operator that represents the boundary interaction 
can be identified with $\Phi^{\sigma,\sigma}_{1}(x)$, see \CLQDbranes, \S2.

\item{\bf 2.} The dependence of $\Phi^{\sigma_2\sigma_1}_{\alpha}(x)$
w.r.t. the variable $\al$ is not only {\it meromorphic}, 
it is analytic for $0 \leq\Re(\al)\leq Q$. This means in particular
that correlation functions like 
\eqn\xxx{
\Big\langle \,
\prod_{i=1}^n \Phi^{\sigma_{i+1}\sigma_i}_{\alpha_i}(x_i)\,\Big\rangle
}
are meromorphic w.r.t. each variable $\al_i$, with poles that
can be avoided by a variation of or by smearing over the
remaining variables $\al_j$, $j\neq i$. 

\noindent
While the first of these statements is easily checked, 
it is nontrivial to prove the second. 
The main ingredients of the proof are contained in
the appendices A and C.

\subsec{Analyticity in $\mub$}

Much of the following will rely on the fact that 
boundary Liouville depends {\it analytically} on the boundary parameter
$\si$ introduced in \smub. In order to exhibit the analytic properties
w.r.t. $\si$ more clearly let us introduce
yet another class of boundary fields as
\eqn\bdrenormtwo{
\tilde{\Phi}^{\sigma_2\sigma_1}_{\alpha}(x)\;=\;G_{\al}^{\si_2\si_1} 
\Psi^{\sigma_2\sigma_1}_{\alpha}(x),
}
where $G_{\al}^{\si_2\si_1}$ is related to $g_{\al}^{\si_2\si_1}$
by canceling the factors that depend on $\si_1$, $\si_2$ only, 
\eqn\bigGsmallg{
G_{\al}^{\si_2\si_1}\;=\;\frac{\Ga_b^2(Q)}
                              {\Ga_b(2\si_1)\Ga_b(2Q-2\si_2)}
g_{\al}^{\si_2\si_1}\;.
}
The analytic properties of the corresponding structure 
functions are analyzed
in the appendices A-C. These results imply that 
$$
\mathboxit{
\eqalign{
{}& {\rm The\;correlation\; functions}\cr
{}& \big\langle\,
V_{\al_n}(z_n,\bz_n)\dots V_{\al_1}(z_1,\bz_1)
\tilde{\Phi}^{\sigma_1\sigma_m}_{\beta_m}(x_m)\dots
\tilde{\Phi}^{\sigma_2\sigma_1}_{\beta_1}(x_1)
\,\big\rangle \cr 
{}& {\rm are\; entire\; analytic\; w.r.t.\;\;} \si_1,\dots,\si_m.
}}
$$
With the help of eqn. \smub\ one may translate 
analyticity w.r.t. the 
boundary parameter $\sigma$ 
back into a statement about the analytic properties
of boundary Liouville theory w.r.t. $\mub$. 
One should note that the structure functions do not share the 
periodicity of $\mub$ under $\si\rightarrow \si+b^{-1}$, which means that
the theory exhibits {\it monodromies} if one considers the analytic
continuation w.r.t. $\mub$ along closed paths.
This may be described
by introducing 
two branch cuts into the complex $\mub$-plane, one running from $-\infty$ to 
$-\sqrt{\frac{\mu}{\sin\pi b^2}}$ and one between
$-\sqrt{\frac{\mu}{\sin\pi b^2}}$ and  
$\sqrt{\frac{\mu}{\sin\pi b^2}}$. We will later discuss a particularly
interesting example of such a monodromy.

\subsec{The spectrum II - Bound states}

We shall now discuss the case where $\si\in\QR$. To simplify the 
discussion slightly we will focus on the cases where the 
boundary conditions $\si_2$ and $\si_1$ 
to the left and right of the insertion point of a 
boundary operator are always equal, $\si_2=\si_1\equiv \si$. 
The corresponding boundary operators will 
be denoted as $\Psi^{\sigma}_{\al}(x)$. 
We will furthermore impose the 
``Seiberg-bound'' $\si<Q/2$ throughout. 

The spectrum now shows an interesting dependence w.r.t. $\si$. 
It remains unchanged as long as $\frac{Q}{4}<\si<\frac{Q}{2}$.
Of primary interest for us will be the case that 
$\si<\frac{Q}{4}$, where one finds in addition to \spec\ 
a discrete part in the spectrum \TLbound:
\eqn\discspec{
\CH_{\sigma\sigma}^{\rm B}\;=\;
\int\limits_{\QS}d\alpha\; \CV_{\alpha,c}\;\oplus\;
\bigoplus_{\alpha\in\QD_s}\CV_{\alpha,c}\;,
}
where 
$\QD_s=\{\alpha\in\QC;\alpha=2\si+nb+mb^{-1}\,,\,n,m\in\QZ^{\geq 0}\}$. 
For $\si<0$ one finds non-unitary representations of the Virasoro 
algebra, which is why we will mostly discuss $0\leq\si<\frac{Q}{4}$ 
in the following. In the case of our main interest, $b=1$, 
we find only a single bound state at $\alpha=2\sigma$ as long as $\si>0$.

In order to get an alternative point of view on the origin of a 
discrete part in the spectrum let us note that the operators 
$\Phi^{\si}_{\al}(x)$ create states $w_{\alpha;\sigma}$
with asymptotic 
wave-functions $\phi_{\alpha}^{\si}(q)$, 
$q:= \int_{0}^{\pi}dx\varphi(x)$,
such that
\eqn\wvfasym{
\phi_{\alpha}^{\si}(q)\;\sim\;
\frac{1}{\sqrt{2\pi}}\, e^{-\frac{Q}{2}q}\big(\, g_{\al}^{\si}\,
e^{\alpha q}+g_{Q-\al}^{\si}
\,e^{(Q-\alpha)q}\big)\quad{\rm for}\;\; q\rightarrow -\infty\;,
}
The asymptotics \wvfasym\ implies that the
wave-functions $\phi_{\alpha}^{\si}(q)$ will be 
generically non-normalizable
for $\Im(\alpha)\neq \frac{Q}{2}$. However, the zero of $g_{\al}^{\si}$
at $\alpha=2\si$ suggests that $w_{\si}:= w_{2\si;\sigma}$ 
represents a {\it normalizable} state.

It can indeed be shown directly that the corresponding vertex operator 
$\Phi_{\si}(x):= \Phi_{2\si}^{\si}(x)$ 
creates normalizable states: The two-point-function of the
operator 
$\Phi_{\al}^{\si}(x)$ can be constructed as
\eqn\twoptphidef{
d^{\si}_{\al}\,|x_2-x_1|^{-2\De_{2\si}}:=  \;
\langle \Phi^{\si}_{\al}(x_2)\Phi^{\si}_{\al}(x_1)\rangle
 =\lim_{\alpha_3\ra 0}
\langle\Phi^{\si}_{\alpha_3}(x_3)
\Phi^{\si}_{\al}(x_2)\Phi^{\si}_{\al}(x_1)
\rangle\;.
}
This allows us to calculate $d_{\al}^{\si}$ from the 
OPE-coefficients $C^{\si_3\si_2\si_1}_{\alpha_3\alpha_2\alpha_1}$.
The result is infinite (proportional to $\de(0)$) for $\al\neq 2\si$, but
turns out to be 
% if we take into account that the three point function 
% $C^{\si_3\si_2\si_1}_{\alpha_3\alpha_2\alpha_1}$ is related to  
% $C^{\si_3\si_2\si_1}_{\alpha_3|\alpha_2\alpha_1}$ via
% $C^{\si_3\si_2\si_1}_{\alpha_3\alpha_2\alpha_1}=
% C^{\si_3\si_2\si_1}_{Q-\alpha_3|\alpha_2\alpha_1}$.
{\it finite} if $\al=2\si$. It is then given by the expression
\eqn\twoptphi{
\eqalign{
d_{\si}\,:=\,& \bigl(\pi \mu \gamma(b^2) 
b^{2-2b^2}\bigr)^{\frac{Q}{2b}}
\frac{\Ga_b^{}(Q-4\si)\Ga_b^2(2Q-2\si)}{\Ga_b^{}(2Q-4\si)\Ga_b^2(Q-2\si)}
\cr  =\;& 2\pi
\bigl(\pi \mu \gamma(b^2)\bigr)^{\frac{Q}{2b}}\frac{Q-4\si}{(Q-2\si)^2}
\frac{\Ga\big(b(Q-4\si)\big)\Ga\big(b^{-1}(Q-4\si)\big)}
     {\Ga^2\big(b(Q-2\si)\big)\Ga^2\big(b^{-1}(Q-2\si)\big)}    .
}}

Let us furthermore note that the operator product expansion
\OPE\ has an analytic continuation to more general values of 
$\si_1,\si_2,\si_3$ and $\al_1,\al_2$, see the Appendix C for 
a more detailed discussion. Here we are interested in the
case that $\si_i=\si<\frac{Q}{4}$, $i=1,2,3$. The first 
important point to observe is that
in the analytically continued OPE one will always find a 
discrete contribution proportional to $\Phi_{\si}$, which was to
be expected due to the appearance of $w_{\si}$ in the
discrete part of the spectrum $\CH^{\rm B}_{\si,\si}$.

In the case that will be our main focus later, namely
$2b^2>1$ and $Q<2\Re(\al_1+\al_2)<2Q$, it turns out that 
the contribution proportional to $\Phi_{\si}$ is the only one that
appears discretely. The operator
product expansion for $\Phi^{\sigma}_{\al_2}(x_2) 
\Phi^{\sigma}_{\al_1}(x_1)$
then takes the following form:
\eqn\phiOPE{
\eqalign{
\Phi^{\sigma}_{\al_2}(x_2) \Phi^{\sigma}_{\al_1}(x_1)\;=\; 
\int_{\QS}d\alpha_3\; & E^{\si}_{\al_3|\al_2\al_1}
\,|x_2-x_1|^{\De_{\alpha_3}-\De_{\al_2}-\De_{\al_1}}\,
\Phi^{\sigma}_{\alpha_3}(x_1)\cr
+ & E^{\si}_{\al_2\al_1}
|x_2-x_1|^{-\De_{2\si}}\Phi_{2\si}^{\si}(x_1)+\;{\rm descendants}.
}}
Of particular importance for us will be the operator 
product coefficient
$E_{\si}:= E^{\si}_{2\si,2\si}$, which has the explicit expression
\eqn\bdstope{
E_{\si}=\frac{\Ga_b(2Q-6\si)}{\Ga_b(Q)}
\frac{\Ga_b(2Q-2\si)}{\Ga_b(Q-4\si)}
\frac{\Ga_b^2(Q-2\si)}{\Ga_b^2(2Q-4\si)}\;.
}
The derivation of this expression is given in the Appendix A.4.

\subsec{The limit $\si\ra 0$}

A particularly interesting value for the boundary parameter $\si$ 
turns out to be $\si=0$. 
To begin with, let us note \MartinecKA\ that 
\eqn\bdstdecomp{
\langle B_{0}|\;=\;\langle B_{b}|+\langle B_{\rm D}|\;,
}
where $\langle B_{\rm D}|=\int dP\,\Psi_{ZZ}\langle P|$ is the 
boundary state introduced in \ZZstate. 
Let us next note that the spectrum of boundary Liouville theory
at $\si=0$ contains, in particular, the vacuum representation 
$\CV_{0,c}$, as follows either from the Cardy-type computation
in \TLbound, or by using the decomposition \bdstdecomp\ together 
with formula (5.9) from \ZZB.

These observations suggest that one may 
view the boundary Liouville theory with $\si=0$ as a kind of 
superposition of the boundary Liouville theory with $\si=b$ and the theory 
corresponding to the Dirichlet type boundary conditions from 
\ZZB. This expectation turns out to be realized in a rather
accurate sense. In the rest of this section we will
summarize some interesting features of boundary Liouville theory 
at $\si=0$ which are derived in the Appendix D.

The vertex operator $\Phi_0$ that corresponds to the highest weight
state in $\CV_{0,c}$
can be constructed by taking the limit $\si\ra 0$ of the 
boundary field $\Phi_{\si}(x)$. It is shown in Appendix D that 
\eqn\zerolimone{
\mathboxit{
\Phi_0\;=\;\lim_{\si\ra 0}\Phi_{\si}(x)\;=\;\Pi_0,
}}
where $\Pi_0$ denotes the {\it projection} onto the 
subspace $\CV_{0,c}$ in $\CH^{\rm B}_{0,0}$.

In order to construct the fields that create the states in the 
complement $\CV_{0,c}^{\perp}$
of $\CV_{0,c}^{}$ one needs to consider the 
boundary fields $\tilde{\Phi}^{\si\si}_{\al}(x)$ instead of 
$\Phi^{\si\si}_{\al}(x)$, cf. \bdrenormtwo. We define
$ \tilde{\Phi}_{\al}(x):=\lim_{\si\ra 0}\tilde{\Phi}^{\si\si}_{\al}(x)$.
The sectors $\CV_{0,c}^{}$ and $\CV_{0,c}^{\perp}$ turn out to be
completely decoupled, in the sense that all mixed correlation functions
which contain fields of both types $\Phi_0$ and $\tilde{\Phi}_{\al}$ 
vanish.

The boundary field $\Phi_0$ acts as a projector in yet another way.
We have 
\eqn\projprop{
\mathboxit{
\eqalign{
\big\langle\,B_{0}\,|\,\Phi_0\,
V_{\al_n}(z_n,\bz_n) & \dots V_{\al_1}(z_1,\bz_1)
\,|\,0\,\big\rangle\;=\cr
 & \;=\; \big\langle\,B_{\rm D}\,|\,
V_{\al_n}(z_n,\bz_n)\dots V_{\al_1}(z_1,\bz_1)
\,|\,0\,\big\rangle.
}}
}
This means that inserting $\Phi_0$ on the boundary of a disk with
the $\si=0$ boundary condition projects out the couplings to 
the term $ \langle B_{b}\,|$ in \bdstdecomp.
Equation \projprop\ encodes a rather remarkable relation 
between the two different types of boundary conditions in
Liouville theory.

Let us finally remark that the analytic continuation w.r.t. $\si$
from $\si=b$ to $\si=0$ corresponds to analytically continuing w.r.t.
$\mub$ along a closed cycle starting from
$$
\mub=-\sqrt{\frac{\mu}{\sin\pi b^2}}\cos\pi b^2,
$$
and returning to the same value for $\mub$. Performing the analytic 
continuation along such a monodromy cycle 
creates an admixture represented by $\langle B_{\rm D}|$.

\newsec{Boundary perturbations}

Our aim will be to study the renormalization group flow 
generated by boundary perturbations of the following form
\eqn\pertop{
S_{\rm pert}\;:=\;-\lambda_{\al}a^{-y_{\al}}
\int_{\partial\Sigma}dx \;\Phi^{\si}_{\alpha}(x),
}
where we will mostly assume that the parameter 
$y_{\al}:= 1-\De_{\alpha}$ satisfies $1\gg y_{\al} >0$. 
Our discussion will partially follow the treatment of similar 
problems in \refs{\cardylec,\Kondo,\boundpert}, concentrating onto 
the new features that originate from the continuous 
spectrum of boundary Liouville theory.

\subsec{Boundary renormalization group flows}

The correlation functions in the perturbed theory are 
formally defined by
\eqn\pertcorr{
\langle\,\CO\,\rangle^{}_{\sigma;\lambda_{\al}}\;:=\;
\langle\,\CO\,{\rm Pexp}(-S_{\rm pert})\,\rangle^{}_{\sigma},
}
where $\CO$ represents the operator insertions that define the 
correlation function in question, and ${\rm Pexp}$ 
is the path ordered exponential. 

Turning on the perturbation \pertop\ will of course spoil 
scale invariance. Let us check explicitly that 
$T(z)\neq \bar{T}(\bar{z})$ when $\la\neq 0$. We will temporarily 
adopt the choice $\Sigma=\QH^+$, the upper half plane.
To first order in $\la$ we have to consider
\eqn\scinvviol{
\eqalign{
\lim_{\ep\downarrow 0}\;\int\limits_{\pa\Sigma}dy\,  \Big\langle
\big(T( & x+i\ep) -\bar{T}(x-i\ep)\big)\,
\Phi^{\si}_{\alpha}(y)\;\CO\,\Big\rangle\cr
& =\lim_{\ep\downarrow 0}\;\int\limits_{\pa\Sigma}dy\,
\bigg\langle \bigg[ 
\bigg(\frac{\De_{\alpha}}{(x-y+i\ep)^2}-\frac{\De_{\alpha}}
     {(x-y-i\ep)^2}\bigg)\Phi^{\si}_{\alpha}(x)
+\cr
&\qquad\qquad\qquad\quad +\bigg(\frac{1-\De_{\alpha}}{x-y+i\ep}-
\frac{1-\De_{\alpha}}{x-y-i\ep}\bigg)
\pa_x\Phi^{\si}_{\alpha}(x)\bigg]
\CO\bigg\rangle\cr
&= 2\pi(\De_{\alpha}-1)\big\langle\,\pa_x\Phi^{\si}_{\alpha}(x)\CO\,\big\rangle.
}
}
This can be seen as an analogy to 
the well-known fact that in the case of bulk perturbations 
the trace of the energy-momentum tensor is proportional to the 
perturbing field itself, see e.g. \cardylec, \S 6.1.

A useful tool for describing the scale-dependence of the 
perturbed theory is the renormalization group (RG). 
Let us consider the effective action defined by
choosing $\Sigma=D_L$, a disk with circumference $2\pi L$, and by 
introducing the ultraviolet cut-off $|x_i-x_j|<a$ in the integrals
that one gets in the perturbative expansion of \pertcorr.
In order to make the effective action independent of the choice
of cut-off we will have to compensate the result of a change of $a$
by a corresponding change of the bare coupling constant $\la_{\al}$.
The coupling constant $\la_{\al}$ will thereby become dependent 
on the dimensionless scale-parameter $l=\ln(a/L)$. To first order 
in $\la$ we find from the explicit scale dependence in \pertop\
that a variation $\de_{\ep}:a\rightarrow a(1+\ep)$ of the cut-off
must be compensated by a variation $\de_{\ep}\la_{\al}=y\la_{\al}\ep+
\CO(\la^2_{\al})$.

In order to find the contribution to $\de_{\ep}\la$ of order
$\CO(\la^2)$ we have to calculate
\eqn\latwo{
\eqalign{
\de_{\ep}^{(2)}\;:=\;\de_{\ep} & \left(\frac{1}{2}
\int_{\partial\Sigma}d\varphi_2d\varphi_1  \;
\Phi^{\si}_{\alpha}(\varphi_2)\Phi^{\si}_{\alpha}(\varphi_1)
\Theta\big(|\varphi_2-\varphi_1|-e^l\big)\right)
\cr
& \qquad\qquad\qquad=-\frac{1}{2}\int_{\partial\Sigma}d\varphi_2d\varphi_1  \;
\Phi^{\si}_{\alpha}(\varphi_2)\Phi^{\si}_{\alpha}(\varphi_1)\de\big(
|\varphi_2-\varphi_1|-e^l\big),
}
}
where we have parametrized $\pa D_L$ by coordinates $\varphi_i$, $i=1,2$.
For small $a/L$ we may  use the OPE of 
$\Phi^{\si}_{\alpha}(\varphi_2)\Phi^{\si}_{\alpha}(\varphi_1)$ in order 
to further evaluate  \latwo.
To keep the discussion simple let us assume that $2b^2>1$. 
The condition that $\Phi^{\si}_{\alpha}(x)$
defines a nearly marginal boundary perturbation then  requires that
$\frac{Q}{4}<\al<b$. The OPE \phiOPE\ is applicable under
these conditions and takes the simple form
\eqn\phiOPEtwo{
\Phi^{\sigma}_{\al}(\varphi_2) \Phi^{\sigma}_{\al}(\varphi_1)\;=\; 
E^{\si}_{\al\al}|\varphi_2-\varphi_1|^{\De_{2\si}-2\De(\al)}
\Phi_{2\si}^{\si}(\varphi_1)+\;{\rm irrelevant\; fields}.
}
Inserting \phiOPEtwo\ into \latwo\ yields the term
\eqn\scalevar{
\de_{\ep}^{(2)}\;=\;-\ep \,E^{\si}_{\al\al} \,
\left(\frac{a}{L}\right)^{\De_{2\si}-2\De(\al)}\!
\int_{\partial\Sigma}d\varphi_1\;
\Phi_{\si}(\varphi_1),
}
which has to be cancelled by a corresponding variation of $\la_{\al}$
with the opposite sign.
We clearly have to distinguish two cases:

\item{\bf 1.} Let $\al\neq 2\si$. Eqn. \scalevar\ implies that 
$\la_{\al}$ shows no scale dependence in second order of 
perturbation theory. It is rather clear that this will also 
be found in higher orders of boundary perturbation theory, 
the reason being simply the absence of $\Phi^{\si}_{\al}$ in the OPE 
for $\al\neq 2\si$. On the other hand we find that 
an additional perturbation 
by \pertop\ with $\al=\si$ is generated at $\CO(\la_{\al}^2)$.

\item{\bf 2.} Let $\al=2\si$. We then find that the scale variation of 
$\la:= \la_{2\si}$ is given by
\eqn\RGla{
\de_{\ep}\la\;=\;\ep(y\la+E_{\si}\la^2)+\CO(\la^3), 
\quad{\rm with}\;E_{\si}:= E^{\si}_{2\si,2\si}.
}

We have arrived at a remarkable conclusion: Only the coupling $\la$ 
that corresponds to the field which creates the 
{\it normalizable} state $w_{\si}$
in the boundary Liouville theory
shows a nontrivial scale-dependence, all others are ``frozen''. 

\subsec{Determination of the new fixed point}

The dependence of the coupling $\la$ w.r.t. the scale-parameter
$l=\ln(L/\epsilon)$ is determined by the flow-equation that
follows from \RGla.
Ignoring terms of $\CO(y)$, where $y:= y_{2\si}$ we find
\eqn\RGflow{
\dot{\la}\;:=\;\frac{d\la}{dl}\;=\;\la y+E_{\si}\la^2+
\CO(\la^3), \quad{\rm with}\;E_{\si}:= E^{\si}_{2\si,2\si}
}

For small values of $ y$ one therefore finds a nontrivial 
fixed point of the renormalization group flow, $\be(\la_{\ast})=0$, at
\eqn\fixpt{
\la^{\ast}\;=\;-\frac{ y}{E_{\si}}+
\CO( y^2)\,.
}

For consistency we need to make sure that the condition that 
the perturbation is nearly marginal, $0< y\ll 1$ really implies that
$\la^{\ast}$ is small. It will be convenient to use $\ep:= b-2\si$
as the small parameter from now on.
For $b\neq 1$ one has $ y=\CO(\ep)$,
and $E_{\si}$ is analytic for $2\si$ near $b$, so that $\la^{\ast}=\CO(\ep)$.
In the case $b=1$ we find that $ y=\CO(\ep^2)$, but
$E_{\si}=\CO(\ep)$ due to the factor $1/\Ga_b(2-4\si)$ in \bdstope,
so that again $\la^{\ast}=\CO(\ep)$.

It remains to describe the new fixed point at $\la=\la^{\ast}$ 
in terms of the known boundary conditions labelled by $\si$.
We may observe 
that $\mub$ is the coefficient of the 
operator $\int_{\pa\Sigma}dx \Psi^{\si\si}_b(x)$ 
which creates {\it non-normalizable} states. The perturbing
operator $\int_{\pa\Sigma}dx \Phi_{\si}(x)$, on the other hand,
creates {\it normalizable} states. This strongly indicates that
the theory at the fixed point can not have a value of 
$\mub$ which differs from the unperturbed theory\foot{I would 
like to thank J. Maldacena
for drawing my attention to this fact, which led me to 
correct an important error in a previous version of this paper.}.
The most natural candidate for the boundary parameter 
$\si_{\ast}$ at the new fixed point is therefore
\eqn\newsi{
\si_{\ast}\;=\;\frac{b}{2}+\frac{\ep}{2}\,,\quad 
{\rm given\;\,that}\quad \si\;=\;\frac{b}{2}-\frac{\ep}{2}\,.
}
$\si_{\ast}$ and $\si$
correspond to the same value of $\mub$, but 
parameterize points on different sheets of the 
branched covering of the $\mub$-plane that boundary Liouville 
theory lives on.

In order to verify that \newsi\ indeed holds, let
us consider the leading order correction 
$\de A^{\si}_{\al}$ to the 
one point function $A^{\si}_{\al}$ which is given by the expression
\eqn\oneptcorr{
\eqalign{
\de A^{\si}_{\al}\;=\;&
\la^{\ast}a^{-y}\int_{\pa\Sigma}dx\;
\langle B_{\si}|\,\Phi_{\si}(x)
V_{\al}(0)\,|\,0\,
\rangle\, \cr
\;=\;&2\pi\;\la^{\ast}\;
\langle B_{\frac{b}{2}}|\,\Phi_{\frac{b}{2}}(1)
V_{\al}(0)\,|\,0\,
\rangle\, +\CO(\ep^2).
}
}
The explicit expression for the bulk-boundary two-point function
in \oneptcorr\ is derived in Appendix B.3. 
It is given by the expression
\eqn\hosopert{
\langle   B_{\frac{b}{2}}|\,
\Phi_{\frac{b}{2}}(1)V_{\al}(0)\,|\,0\,
\rangle
\;=\;\frac{1}{\sqrt{2\pi}}\frac{2\pi bP}{\sinh 2\pi bP}
\Psi_{\rm ZZ}(P)\,.
}
Let us furthermore observe that the expression for $\la^{\ast}$
may be simplified as
\eqn\lambdastern{
\la^{\ast}\;=\;-y E_\si ^{-1}\;=\;-\ep b^{-1}+\CO(\ep^2)\, .
}
By inserting \hosopert\ and \lambdastern\ into
\oneptcorr\ we finally arrive at the expression
\eqn\oneptcorrfin{
\de A^{\si}_{\al}\;=\;\ep \,\sqrt{\frac{\pi}{2}}\,
\frac{-4\pi P}{\sinh 2\pi bP}\Psi_{\rm ZZ}(P)\,.
}
In order to verify \newsi\ it remains to observe that
$A^{\si}_{\al}+\de A^{\si}_{\al}$ as given by \oneptcorrfin\ may
also be written as
\eqn\oneptcorralt{
A^{\si}_{\al}+\de A^{\si}_{\al}\;=\;A^{\si}_{\al}+\ep \,\left[
\frac{\pa}{\pa\si}A_{\al}^{\si}\right]_{\si=\frac{b}{2}}\;=\;
A^{\si_{\ast}}_{\al}+\CO(\ep^2)\,.
}
 
\subsec{A simple, but interesting RG flow}

There is an interesting example of a boundary perturbation
for which the theory remains exactly soluble.
Let us consider the boundary Liouville theory at $\si=0$.
In \S2.6 we had observed that the primary boundary field $\Phi_0$
projects $\CH_{0,0}^{\rm B}$ onto the sector $\CV_{0,c}$, and 
$\langle B_{0}|$ onto $\langle B_{\rm D}|$. These properties will
imply that the RG flow generated by the boundary perturbation $\Phi_0$
is almost trivial. Let us consider 
\eqn\decone{
\eqalign{
\big\langle B_{0,\la}|\,V_{\al}(z,\bz)\,\big\rangle\;:=\;&
\big\langle B_{0}|\,{\rm Pexp}\Big(-\frac{\la}{a}\int_{\partial\Sigma}dx
\,\Phi_0\Big)
V_{\al}(z,\bz)\,\big\rangle\cr
\;=\;&
\big\langle B_{0}|\,{\rm exp}\Big(-2\pi\la 
\frac{L}{a}\Phi_0\Big)
V_{\al}(z,\bz)\,\big\rangle
}}
By using $\Phi_0^2=\Phi_0$ one may further calculate
\eqn\dectwo{
\big\langle B_{0,\la}|\,V_{\al}(z,\bz)\,\big\rangle\;=\;
\big\langle B_{0}|\,\big(1+(
e^{-2\pi\la\frac{L}{a}}-1)\Phi_0\big)V_{\al}(z,\bz)\,\big\rangle.
}
The endpoint of the flow is therefore represented by 
\eqn\decthree{
\eqalign{
\langle B_{0,\infty}|\,V_{\al}(z,\bz)\,\big\rangle\;=\;&
\big\langle B_{0}|\,(1-\Phi_0)V_{\al}(z,\bz)\,\big\rangle\cr
\;=\;&\big\langle B_{b}|\,V_{\al}(z,\bz)\,\big\rangle,
}}
where we have used equations \projprop\ and \bdstdecomp\
to go from the first to the 
second line. This means that the RG flow generated by the boundary
perturbation $\Phi_0$ acts again like a covering transformation: 
It removes the admixture represented by $\langle B_{\rm D}|$ 
that is created by performing the
analytic continuation w.r.t. $\mub$ along a closed path 
starting from $\mub=-\sqrt{\frac{\mu}{\sin\pi b^2}}\cos\pi b^2$.

\newsec{D-brane decay in two-dimensional string theory}

Let us consider D1-branes in noncritical string theories that are 
characterized by boundary states of the form
\eqn\bdst{
(B|\;=\;\langle B_{\rm N}|_{X^0}^{}\ot\langle B_{\si}|,
}
where $\langle B_{\rm N}|_{X^0}^{}$ 
is supposed to described Neumann type boundary 
conditions for the $X^0$-CFT to begin with. We will assume 
that the reader is familiar with the discussion in \S3 of 
\CLQDbranes. Throughout this section we will consider the case
of $b=1\Leftrightarrow c=25$.

In the two cases $\si\in\QS$ and $\si\in\big(\frac{1}{2},1\big)$,
the Liouville theory on the strip with boundary conditions labelled 
by $\si$ on both sides will have a purely continuous spectrum given by
\spec. This implies a purely continuous spectrum of 
open string tachyons on the D1-branes, which is generated by the 
on-shell tachyon vertex operators
\eqn\tachyonvert{
T_{E}\;=\;\big[ e^{i\omega X^0}
\Phi^{\si}_{\al}\big]^{}_{\al=1+i\omega}.
}
The situation changes when $\si$ gets smaller than $\frac{1}{2}$.
The bound states in $\CH_{\sigma_2\sigma_1}^{\rm B}$, \cf\ 
\discspec,
then yield on-shell tachyon vertex operators with imaginary
frequencies like
\eqn\realtachyonvert{
T_{E}\;=\;\big[e^{\nu X^0}
\Phi^{\si}_{2\si}\big]^{}_{2\si=1-\nu}.
}
Appearance of imaginary energies usually signals some instability.
The very form of \realtachyonvert\ suggests that a perturbation of the
system by \realtachyonvert\ has an effect that blows up exponentially
when time $X^0\ra\infty$. 

% Appearance of modes with 
% imaginary energy is furthermore well-known to produce 
% divergencies in loop amplitudes. 

% Unfortunately we do not know an exact boundary CFT that would correspond
% to a boundary perturbation of the form \realtachyonvert.
% So it seems that the best we can do is to interpret 
% our previous results on boundary RG flows in Liouville theory in 
% terms of decay processes of the D1-branes in noncritical string theories.

\subsec{Comments on the time-dependent description}

Ideally we would like to associate
time-dependent solutions of noncritical open string theory to
the RG flows discussed in the previous section. 
The basic idea for a perturbative construction of a 
time-dependent solution is rather obvious:
Replace the couplings of the relevant perturbing fields 
by (yet undetermined) functions 
of the time coordinate $X^0$. These functions have to be chosen 
such that conformal symmetry is preserved by the resulting action.
To begin with, let us observe that an ansatz of the simple form
\eqn\deltaS{
\de S\,=\,\kappa\int_{\pa\Sigma}dx \;\big(e^{\nu X^0}
\Phi^{\si}_{1-\nu}\big)(x)
}
will not preserve conformal invariance in general. At second order in 
$\kappa$ one has to consider
\eqn\cfviol{
\frac{1}{2}\kappa^2 \;{\rm P}\!\!\int_{\pa\Sigma}dx_2dx_1\;
\big(e^{\nu X^0}\Phi_{\al}^{\si}\big)(x_2)
\big(e^{\nu X^0}\Phi_{\al}^{\si}\big)(x_1)\;.
}
We will again focus on the case that 
$\Phi_{\al}^{\si}(x)$ is nearly marginal, $0<\nu^2\ll 1$.
The main contribution to the integral
\cfviol\ then comes from the vicinity of the diagonal $x_2=x_1$, and
may again be estimated by using the operator product 
expansion. We thereby find a conformally non-invariant
contribution of the form
\eqn\cfvioltwo{
\frac{1}{2}\kappa^2\, \frac{c_{\nu}}{\nu^2}\,
E_{\al\al}^{\si} 
\int_{\pa\Sigma}dx\;
\big(e^{2\nu X^0}\Phi_{\si}\big)(x)+\CO(\nu^0),\;
}
where $c_{\nu}$ is constant up to terms of higher order in $\nu^2$.
We must therefore modify our ansatz \deltaS\ 
by corrections of higher order in $\kappa$ which
will contain the macroscopic boundary field $\Psi_{\si}(x)$
multiplied by $e^{2\nu X^0}$. Note that the first order term
\deltaS\ is the leading one for $X^0\ra -\infty$, however.

One may then try to construct $\de S$ in the form
\eqn\deltaS{
\de S\,=\,\kappa\int_{\pa\Sigma}dx \;\big(e^{\nu X^0}
\Phi^{\si}_{1-\nu}\big)(x)+\sum_{n=2}^{\infty}\kappa^n
\,(\de S)^{(n)}[X^0].
}  
It seems clear that the higher order 
corrections in $\de S$ are also
governed by the OPE of Liouville boundary
fields. We had previously observed that the OPE of the 
boundary fields 
$\Phi_{\al}^{\si}(x_2)\Phi_{\al}^{\si}(x_1)$
will generically not contain the field 
$\Phi_{\al}^{\si}(x_1)$ at all (unless $\al=2\si$). 
In this case we would consequently
expect that the higher order corrections $(\de S)^{(n)}$ 
do not contain $\Phi_{\al}^{\si}(x_1)$ as well.
This phenomenon can be seen as a counterpart of the 
``frozen'' couplings that we had encountered in the previous section.

\subsec{Unstable vs. ``frozen'' modes}

In either picture we observe an important dichotomy. Perturbations
containing microscopic boundary fields generate trivial RG
flows / carry purely exponential time dependence, unlike 
the perturbations which contain macroscopic boundary fields.
The crucial difference between these two types of perturbations
originates from the different character of the states that 
are created from the two types of perturbations:
The macroscopic boundary fields create {\it normalizable} states, 
as opposed to the microscopic boundary fields.

One should observe, however, that the two cases are fundamentally 
different also from the space-time point of view. The
microscopic boundary fields describe open string tachyon 
configurations which diverge in the weak coupling region
$\varphi\ra -\infty$, as opposed to the case of 
macroscopic boundary fields. Only the 
normalizable tachyon field configurations 
will appear in quantum fluctuations.
This leads us to propose that 
the perturbations that contain macroscopic boundary fields 
describe brane decays that can take place spontaneously, whereas
perturbations by microscopic boundary fields describe 
decays triggered by external sources instead.
A similar proposal about 
perturbations by bulk fields 
was made some years ago by Seiberg and Shenker \SeiShen. 
Our previous discussion exhibits the world-sheet
origin of this phenomenon in a prototypical example. 

\subsec{A fast decay}

The perturbative treatment of time-dependent phenomena is clearly
limited to slow processes or to the initial stages of a decay process.
In the present case we only know one example where we can go beyond this
restriction to study a fast decay of a D-brane. 

Let us consider the D-brane described by the boundary state 
\eqn\bdsttwo{
(B_0|\;=\;\langle B_{\rm N}|_{X^0}^{}\ot\langle B_0|,
}
which corresponds to the limit $\si\ra 0$ discussed in \S2.6.
The boundary state $(B_0|$ 
describes a bound state formed by 
a D1 brane with $\murb=\sqrt{\mu_{\rm ren}}$ and a D0 brane. This bound state 
is perturbatively unstable as follows from the existence of a normalizable
highest weight state within $\CH^{\rm B}_{0,0}$.
The corresponding boundary perturbation is obtained by
dressing the macroscopic boundary field $\Phi_0$ with $e^{X^0}$, 
\eqn\dressedphinull{
S_{\rm B}\,=\,\la\int d\tau\,\Phi_0\,e^{X^0}.
}
By means of a calculation that is very similar to the one 
in \S3.3 one finds that the boundary state which describes the 
decay of the brane characterized by $(B_0|$ is given as
\eqn\bdsttwo{
(B_{0,\la}|\;=\;
\langle B_{\rm N}|_{X^0}^{}\ot\langle B_1|+
\langle B_{\rm S,\la}|_{X^0}^{}\ot\langle B_{\rm D}|,
}
where $\langle B_{\rm S,\la}|_{X^0}^{}$ is the relevant member of
the class of boundary states for the $X^0$-CFT 
introduced by A. Sen in \Senone, 
see e.g. \refs{\GutperleXF,\LLM} for further
results and references. The interpretation of the
time-dependent process described by $(B_{0,\la}|$ should be clear:
The D0 brane decays spontaneously, leaving behind the D1 brane with 
$\murb=\sqrt{\mu_{\rm ren}}$ as stable remnant.\foot{In my talk 
at Strings 2003 I have made an incorrect statement about this 
point. I thank J. Maldacena for pointing out that he expects
the scenario above to be realized instead.}

\subsec{Summary: Dependence with respect to $\murb$}

Let us summarize the resulting picture of the dependence 
of the D1 branes w.r.t. the parameter  $\murb$ or $\si$.
We will consider the analytic continuation starting from the case 
that $\sigma\in\QS$ which corresponds to
$\murb>\sqrt{\mu_{\rm ren}}$. One is then dealing with a 
D-string that stretches along the Liouville-direction, 
but gradually disappears in the strong coupling region 
$\varphi\ra\infty$. This follows from the fact that
the one-point function $\langle B_{\si}|P\rangle$ 
decays exponentially as 
\eqn\expdec{
\langle B_{\si}|P\rangle\;\sim\;
e^{-2\pi P(2\Re(\si)-1)}\quad{\rm for}\;\; P\ra\infty\,,
} 
taking into account
that wave-packets 
with average energies $\bar{E}\sim P$
probe more and more deeply into the region $\varphi\ra\infty$
if one increases $\bar{E}$. This means that the
effective coupling between closed string wave-packets 
and the D1 branes with $\si<\frac{1}{2}$ decreases fast
with the depth of penetration into the strong coupling region.
The boundary cosmological constant
$\murb$ determines how far the D1 brane extends into the
strong coupling region: It stretches further out to $\varphi\ra\infty$ 
if one decreases $\murb$. 

The properties of the D1 brane change qualitatively as soon as 
one passes the turning point $\si=\frac{1}{2}\Rightarrow
\murb=-\sqrt{\mu_{\rm ren}}$
upon continuing from $\si>\frac{1}{2}$ to $\si<\frac{1}{2}$:
The D1 brane acquires a ``mass'' near $\varphi\ra\infty$.
This follows from the exponential growth \expdec\ 
of the one-point function $\langle B_{\si}|P\rangle$
by similar arguments as used in the previous paragraph.
The bound state that appears in $\CH_{\si,\si}^{\rm B}$ for 
$\si<\frac{1}{2}$ is naturally interpreted as an 
open string bound to the part of the D1 brane that
is localized near $\varphi\ra\infty$. These branes, however,
are unstable. Near $\si=\frac{1}{2}$ one may 
use the results from \S3.1 and \S3.2 to conclude that
the decay takes place slowly and produces a D1 brane with 
reduced ``mass'' near $\varphi\ra\infty$. 

When $\si$ decreases
further we do not expect qualitative changes of the picture 
as long as we have $\si>0$. At $\si=0$, however, we not only find
that the mass near $\varphi\ra\infty$ becomes {\it equal} 
to the mass of the D0 brane. The results of \S2.6 concerning 
the decoupling of the sector $\CV_{0,c}$ from the rest of 
the open string spectrum demonstrate that there are no open strings
that stretch between the D0 component  of the D1 brane with 
$\si=0$ and the rest.
The D0 part of the D1 brane becomes free to decay independently of the
rest, as discussed in \S3.3, \S4.3.

\bigskip
\centerline{\bf{Acknowledgements}}
We would like to thank V. Kazakov, I. Kostov, 
J. Maldacena, V. Schomerus, N. Seiberg and H. Verlinde
for discussions, comments and interest in the present work.
We furthermore acknowledge support by the SFB 288 of the DFG, 
as well as partial support by the EC via the EUCLID
research training network contract HPRN-CT-2002-00325.
\vfill
\eject

\appendix{A}{Analytic properties of the structure functions}

\subsec{Special functions}

The function $\Ga_b(x)$ is a close relative of the double
Gamma function studied in \Barnes\Shintani. It 
can be defined by means of the integral representation
\eqn\doublegamma{
\log\Ga_b(x)\;=\;\int\limits_0^{\infty}\frac{dt}{t}
\biggl(\frac{e^{-xt}-e^{-Qt/2}}{(1-e^{-bt})(1-e^{-t/b})}-
\frac{(Q-2x)^2}{8e^t}-\frac{Q-2x}{t}\biggl)\;\;.
}
Important properties of $\Ga_b(x)$ are
\item{(i)} {\it Functional equation:} 
$\Ga_b(x+b)=\sqrt{2\pi}b^{bx-\frac{1}{2}}\Ga^{-1}(bx)\Ga_b(x)$.
\item{(ii)} {\it  Analyticity:}
$\Ga_b(x)$ is meromorphic,
 poles:
$x=-nb-mb^{-1}$, $n,m\in\BZ^{\geq 0}$.
\item{(iii)} {\it Self-duality:} $\Ga_b(x)=\Ga_{1/b}(x)$. 

\noindent
The function $S_b(x)$ may be defined by 
\eqn\sbint{
\log S_b(x)\;=\;\int\limits_0^{\infty}\frac{dt}{t}
\biggl(\frac{\sinh t(Q-2x)}{2\sinh bt\sinh b^{-1}t}-
\frac{Q-2x}{2t}\biggl)\;\;.
}
$S_b(x)$ is related to  
$\Ga_b(x)$ via
\eqn\sbdef{
S_b(x)\;=\;\Ga_b(x)/\Ga_b(Q-x)\;.
}
%This function, or close relatives of it like 
%$e_b(x)\;=\;e^{\frac{\pi i}{2}x^2}\,e^{-\frac{\pi i}{24}(2-Q^2)}s_b(x)$,
%have appeared in the literature under various names like 
%``Quantum Dilogarithm'' \cite{FK1}, ``Hyperbolic G-function''
%\cite{Ru}, ``Quantum Exponential Function'' \cite{W} 
%and ``Double Sine Function'', we refer to 
%the appendix of \cite{KLS} for
% a useful collection of properties of $s_b(x)$ and further references.
The most important properties for our purposes are 
\item{(i)} {\it Functional equation:} $S_b(x+b)=
2\sin \pi b x\;
S_b(x)$. 
\item{(ii)} {\it Analyticity:}
$S_b(x)$ is meromorphic,
\eqn\sbanal{ 
\eqalign{{}& {\rm poles:}\;\,  
x=-(nb+mb^{-1}), n,m\in\BZ^{\geq 0}.\cr
{}& {\rm zeros:}\;\,  
x=Q+(nb+mb^{-1}), n,m\in\BZ^{\geq 0}.
}}
\item{(iii)}{\it Asymptotics:}
\eqn\sbas{
S_b(x)=e^{\mp \frac{\pi i}{2}x(x-Q)}\quad
{\rm for}\;\,\Im(x)\ra\pm\infty.
} 
\item{(iv)} {\it Inversion relation:} $S_b(x)S_b(Q-x)\;=\;1.$
\item{(v)}{\it Residue:} 
\eqn\sbres{
{\rm res}_{x=0}S_b(x)
\;=\;(2\pi)^{-1}\,.
}

\subsec{Three point functions}

To begin with, let us quote the explicit formula for 
$C_{\alpha_{3}|\alpha_{2}\alpha_{1}}^{\sigma_{3}\sigma_{2}\sigma_{1}}$
found in \PonTe:
\eqn\OPEcoeffexpl{
\eqalign{ 
{} C_{\alpha_{3}|\alpha_{2}\alpha_{1}}^{\sigma_{3}\sigma_{2}\sigma_{1}} 
= &  \bigl(\pi \mu \gamma(b^2) 
b^{2-2b^2}\bigr)^{\frac{1}{2b}(\alpha_3-\alpha_2-\alpha_1)}
\Gamma_b(2Q-\alpha_1-\alpha_2-\alpha_3)\cr 
& \times \frac{\Gamma_b(\alpha_2+\alpha_3-\alpha_1)
 \Gamma_b(Q+\alpha_2-\alpha_1-\alpha_3)\Gamma_b(Q+\alpha_3-\alpha_1-\alpha_2)}
{\Gamma_b(2\alpha_3-Q)\Gamma_b(Q-2\alpha_2)\Gamma_b(Q-2\alpha_1)\Gamma_b(Q)}\cr
&   \times\frac{S_b(\alpha_3+\sigma_1-\sigma_3)S_b(Q+\alpha_3-\sigma_3-\sigma_1)}
     {S_b(\alpha_2+\sigma_2-\sigma_3)S_b(Q+\alpha_2-\sigma_3-\sigma_2)}  
   \int\limits_{\QR+i0}ds \;\prod_{k=1}^{4}
    \frac{S_b(U_k+is)}
         {S_b(V_k+is)}, 
}}
The coefficients $U_k$, $V_k$ and $k=1,\ldots,4$ are defined as
$$
\eqalign{ 
 U_1 =& \sigma_1+\sigma_2-\alpha_1 , \qquad   
 V_1 =\sigma_2+\bar{\sigma}_3-\alpha_1+\alpha_3, \cr
 U_2 =& \bar{\sigma}_1+\sigma_{2}-\alpha_1,  \qquad    
 V_2 =\sigma_2+\bar{\sigma}_3-\alpha_1+\bar{\alpha}_3, \cr
 U_3 =& \alpha_2+\sigma_2-\sigma_3, \qquad             
 V_3 = 2\sigma_2, \cr
 U_4 =& \bar{\alpha}_2+\sigma_2-\sigma_3,\qquad V_4=Q\;.         } 
$$
We have used the notation $\bar{\sigma}_l:= Q-\sigma_l$, 
$\bar{\alpha}_l:= Q-\alpha_l$, $l=1,2,3$.

Our aim is to study the analytic properties of the three point function
$D_{\alpha_{3}\alpha_{2}\alpha_{1}}^{\sigma_{3}\sigma_{2}\sigma_{1}}$
of the fields $\Phi^{\sigma_2\sigma_1}_{\alpha}(x)$, which can be 
written in terms of 
$C_{\alpha_{3}|\alpha_{2}\alpha_{1}}^{\sigma_{3}\sigma_{2}\sigma_{1}}$
as 
\eqn\CDrel{
D_{\alpha_{3}\alpha_{2}\alpha_{1}}^{\sigma_{3}\sigma_{2}\sigma_{1}}
\;=\;g_{\al_3}^{\si_1\si_3}g_{\al_2}^{\si_3\si_2}g_{\al_1}^{\si_2\si_1}
C_{Q-\alpha_{3}|\alpha_{2}\alpha_{1}}^{\sigma_{3}\sigma_{2}\sigma_{1}}\;.
}
The result is simplest for the three point function 
$D_{\ta_{3}\ta_{2}\ta_{1}}^{\sigma_{3}\sigma_{2}\sigma_{1}}$
of the fields $\tilde{\Phi}^{\sigma_2\sigma_1}_{\alpha}(x)$,
which is related to 
$D_{\alpha_{3}\alpha_{2}\alpha_{1}}^{\sigma_{3}\sigma_{2}\sigma_{1}}$
via \bigGsmallg. We are going to prove the following assertion:
\eqn\analD{
\eqalign{
{}& {\rm The\;dependence\;of}\;
D_{\ta_{3}\ta_{2}\ta_{1}}^{\sigma_{3}\sigma_{2}\sigma_{1}}\;
{\rm is\; meromorphic\; with\; respect\; to\; its}\cr
{}& {\rm six\; variables.}\;\, 
D_{\ta_{3}\ta_{2}\ta_{1}}^{\sigma_{3}\sigma_{2}\sigma_{1}}\;
{\rm has\; poles\; if\; and\; only\; if }\cr 
{}& \qquad\qquad \sum_{k=1}^{3} 
\ep_k(2\al_i-Q)+Q+2(nb+mb^{-1})=0,\cr
{}& {\rm where}\;\ep_k\in\{+,-\}\;{\rm and}\; 
n,m\in\BZ^{\geq 0}.
}}
This set of poles coincides precisely with the set of poles of the 
three point function $C(\al_3,\al_2,\al_1)$ of the primary field
$V_{\al}(z,\bz)$ in the bulk. It is remarkable and important for 
our present aims that the dependence of 
$D_{\ta_{3}\ta_{2}\ta_{1}}^{\sigma_{3}\sigma_{2}\sigma_{1}}$
w.r.t. the variables 
$\si_k$, $k=1,2,3$ is {\it entire analytic}. To establish our claim 
\analD\ becomes straightforward once we understand the meromorphic
continuation of the integral that appears in \OPEcoeffexpl.

\subsec{Meromorphic continuation of certain integrals}

We are forced to study the dependence of the integral
\eqn\integral{
\int\limits_{\QR+i0}ds \;\prod_{k=1}^{4}
    \frac{S_b(U_k+is)}
         {S_b(V_k+is)}
}
on its parameters. The integrand behaves for large $|s|$ as 
$e^{\pi Q|s|\sum_{k=1}^4(U_k-V_k)}$. By noting that 
$\sum_{k=1}^4(U_k-V_k)=-Q$ in our case we may conclude that the 
convergence of the integral \integral\ does not pose any problems.
Let us furthermore note that the integrand of \integral\
has poles at $is=-U_k-nb-mb^{-1}$,
$k=1,\dots,4$, $n,m\in\BZ^{\geq 0}$,
and poles at
$is=Q-V_k+nb+mb^{-1}$, $k=1,\dots,4$, $n,m\in\BZ^{\geq 0}$, in the 
right half plane. As long as $0<\Re(U_k)$ we therefore find all
poles at $is=-U_k-nb-mb^{-1}$ strictly in the left half plane and
if furthermore $\Re(V_k)<Q$ then all poles at
$is=Q-V_k+nb+mb^{-1}$ are localized in the right half plane only. 
We conclude that the integral \integral\ is 
analytic w.r.t. $U_k$ and $V_k$, $k=1,\dots,4$ as long as 
$0<\Re(U_k)$ and $\Re(V_k)<Q$ hold for $k=1,\dots,4$. 

An analytic continuation of the integral \integral\ to 
{\it generic} values of $U_k$, $V_k$ can be defined by replacing the
contour $\QR+i0$ in \integral\ by a contour $\CC$ that is suitably indented
around the strings of poles at $is=-U_k-nb-mb^{-1}$ that have entered 
the right half-plane, as well as the strings of poles
at $is=Q-V_k+nb+mb^{-1}$ that can be found in the left half-plane.
Of course one should require that $\CC$ approaches the real
axis for $|s|\ra\infty$.
Such a contour will exist iff none of the poles at 
$is=-U_k-nb-mb^{-1}$, $k=1,\dots,4$, 
happens to lie on top of a pole at $is=Q-V_l+nb+mb^{-1}$, $l=1,\dots,4$.

Otherwise let us study the behavior of \integral\ 
when $-U_k-nb-mb^{-1}=Q-V_l+n'b+m'b^{-1}+\ep$ with $\ep\downarrow 0$.
The singular behavior of the integral may be determined by 
deforming the contour $\CC$ into a new contour  
that is given as the sum of 
a contour $\CC'$ which separates the pole
at $is=Q-V_l+nb+mb^{-1}$ from $is=+\infty$ and a small circle around 
$is=Q-V_l+nb+mb^{-1}$. The factor $S_b(U_k+s)$ in the 
numerator of the integrand yields a residue proportional to 
$S_b(\ep+(n'-n)b+(m'-m)b^{-1})$ which develops a pole for $\ep\ra 0$
if $n'-n<0$ and $m'-m<0$. This implies that \integral\
has poles if
\eqn\polesintegral{
Q+U_k-V_l+nb+mb^{-1}=0,\quad k,l=1,\dots,4, \quad
n,m\in\BZ^{\geq 0}.
}
One may finally convince oneself that the contour $\CC'$ can always be 
chosen such that integration over $\CC'$ is nonsingular. The list of 
poles given in \polesintegral\ is therefore complete.

\subsec{Special values of the three point functions}

The general expression \OPEcoeffexpl\ simplifies considerably for
certain values of the parameters. Of particular relevance for us 
are the cases where one of $\al_i$, $i=1,2,3$ is set to a value
that parametrizes a discrete representation in the spectrum.
As an example let us consider the case $\al_1=\si_1+\si_2$. The integral 
in  \OPEcoeffexpl\ becomes singular in this case since the 
contour of integration becomes pinched between the poles of 
$S_b(U_1+s)$ and $S_b(V_4+s)$. In order to extract the singular
part of the integral let us deform the contour into the contour 
$\BR-i0$ plus a small circle around $s=0$. Only the contribution 
from the circle around $s=0$ displays a singular behavior for 
$U_1\ra 0$, and is given by
\eqn\residue{
\prod_{k=1}^3\frac{S_b(U_{k+1})}{S_b(V_k)}S_b(U_1).
}
Once we multiply $C_{Q-\alpha_{3}|\alpha_{2}\alpha_{1}}^{\sigma_{3}\sigma_{2}\sigma_{1}}$ by the factor $g_{\al_1}^{\si_2\si_1}$, the 
pole from $S_b(U_1)$ in \residue\ gets cancelled by a zero of
$g_{\al_1}^{\si_2\si_1}$, leading to a finite result for 
$D_{\alpha_{3}\alpha_{2}\alpha_{1}}^{\sigma_{3}\sigma_{2}\sigma_{1}}$.
In this way it becomes straightforward to calculate the 
special values $D^{\si\sigma\sigma}_{\al_3,\al_2,2\si}$, 
$D^{\si}_{\al}:= 
D_{\alpha,2\si,2\si}^{\sigma\sigma\sigma}$ and 
$D_{\si}:= D_{2\si,2\si,2\si}^{\sigma\sigma\sigma}$
explicitly.
\eqn\spOPEcoeffone{
\eqalign{ 
  D^{\si\sigma\sigma}_{\al_3,\al_2,2\si}= 
\bigl(\pi \mu \gamma(b^2) 
b^{2-2b^2}\bigr)^{\frac{Q}{2b}} & 
\frac{\Ga_b(2\si)\Ga_b^3(2Q-2\si)\Ga_b(Q)}
     {\Ga_b(2Q-4\si)\Ga_b(\al_2)\Ga_b(Q-\al_2)\Ga_b(\al_3)\Ga_b(Q-\al_3)}\cr 
\times & \frac{\Ga_b(Q+\al_3-\al_2-2\si)\Ga_b(Q+\al_2-\al_3-2\si)}
        {\Ga_b(Q+\al_3-2\si) \Ga_b(Q+\al_2-2\si)}\cr
\times & \frac{\Ga_b(2Q-\al_3-\al_2-2\si)\Ga_b(\al_3+\al_2-2\si)}
       {\Ga_b(2Q-\al_3-2\si)\Ga_b(2Q-\al_2-2\si)}\,,
}}
\eqn\spOPEcoefftwo{
\eqalign{ 
D^{\si}_{\al}=
\bigl(\pi \mu \gamma(b^2) 
b^{2-2b^2}\bigr)^{\frac{Q}{2b}}& 
\frac{\Ga_b^3(2Q-2\si)}{\Ga_b^2(2Q-4\si)\Ga_b(Q-2\si)}
\cr & \times 
\frac{\Ga_b(Q+\al-4\si)\Ga_b(2Q-\al-4\si)}
     {\Ga_b(Q+\al-2\si)\Ga_b(2Q-2\si-\al)}
}}
and 
\eqn\spOPEcoeffthree{
D_{\si}=
\bigl(\pi \mu \gamma(b^2) 
b^{2-2b^2}\bigr)^{\frac{Q}{2b}}
\frac{\Ga_b^3(2Q-2\si)}{\Ga_b^3(2Q-4\si)}
\frac{\Ga_b(2Q-6\si)}{\Ga_b(Q)}
\;.
}
The epxression for the two-point function 
given in formula \twoptphi\ is obtained from 
\spOPEcoefftwo\ by setting $\alpha=0$. We finally recover the 
formula \bdstope\ for the OPE coefficient $E_{\si}$ 
via $E_{\si}=D_{\si}/d_{\si}$.

\appendix{B}{The bulk-boundary two-point function}

It remains to study the bulk-boundary two-point function 
\eqn\bulkbddef{
A^{\si}_{\be|\al}\;\equiv\;\langle B_{\si}|\,
\Psi^{\si\si}_{\be}(0)V_{\al}(z,\bz)\,|\,0\,
\rangle_{z=\frac{i}{2}}^{}.
}
The following expression for $A^{\si}_{\be|\al}$
was found in \Hosomichi:
\eqn\hosoform{
A^{\si}_{\be|\al}\;=\;N\,
\rho_{\be|\al}\int\limits_{-\infty}^{\infty}dt\;
\prod_{\ep=\pm}\frac{S_b\big(\frac{1}{2}(2\al+\be-Q)+ i\ep t\big)}
     {S_b\big(\frac{1}{2}(2\al-\be+Q)+ i\ep t\big)}\,
e^{2\pi t(2\si-Q)}\,.
}
The prefactor $\rho_{\be|\al}$ in \hosoform\ is defined by
\eqn\hosoformtwo{
\rho_{\be|\al}=
\big(\pi \mu\ga(b^2)b^{2-2b^2}\big)^{\frac{Q-2\al-\be}{2b}}
\frac{\Ga_b^3(Q-\be)\Ga_b(2Q-2\al-\be)\Ga_b(2\al-\be)}
{\Ga_b(Q)\Ga_b(\be)\Ga_b(Q-2\be)\Ga_b(2\al)\Ga_b(Q-2\al)}.
}
The precise numerical value of the normalization factor $N$
is crucial in \S 3.3. We will therefore begin
by describing how to fix the value of $N$.

\subsec{Determination of the normalization factor N}

Since $\lim_{\be\ra 0}\Psi^{\si_2\si_1}_{\al}(x)=1$ \PonTe\
we have to choose 
$N$ such that $A^{\si}_{\be|\al}$ satisfies
\eqn\Anorm{
\lim_{\be\ra 0} A^{\si}_{\be|\al}\;=\; A^{\si}_{\al}\;.
}
In order to evaluate the limit on the left hand side of 
\Anorm\ we need to observe that the integral in \hosoform\
becomes singular for $\be\downarrow 0$ due to a pinching of the 
contour of integration by the following poles:
\eqn\Apoles{
\eqalign{
{\rm UHP:}& \quad a)\;\,2t=i(2\al-Q+\be),
\qquad\quad b)\;\, 2t=i(Q-2\al+\be),\cr 
{\rm LHP:}& \quad c)\;\, 2t=i(2\al-Q-\be),
\qquad\quad d)\;\,2t=i(Q-2\al-\be).
}}
We will deform the contour of integration into a contour that
passes above the poles a) and b) in the upper half plane,
plus two circles around these poles. Taking into account
\sbres\ and \sbdef\ then yields the following formula
\eqn\Anormcalc{
\eqalign{
\lim_{\be\ra 0} A^{\si}_{\be|\al}\;=\;& N\,
\big(\pi \mu\ga(b^2)b^{2-2b^2}\big)^{\frac{Q-2\al}{2b}}
\frac{\Ga_b(2Q-2\al)}{\Ga_b(Q-2\al)}
\frac{2\cos(\pi(2\al-Q)(2\si-Q))}{S_b(2\al)S_b(2Q-2\al)}
\cr
\;=\;& N\,v(P)
\frac{\cosh (2\pi (2\si-Q)P)}
{\sinh 2\pi b P\sinh 2\pi b^{-1}P}\,
}}
where we have set $\al=\frac{Q}{2}+iP$ and used that 
\eqn\valt{
v(P)=\big(\pi \mu\ga(b^2)b^{2-2b^2}\big)^{\frac{Q-2\al}{2b}}
\frac{\Ga_b(2Q-2\al)}{\Ga_b(Q-2\al)}.
}
By comparing with \onept, \ZZstate\ we may now read off
that $N=-2^{\frac{5}{4}}\sqrt{\pi}$.

\subsec{Analyticity of $
\langle B_{\si}|\,
\Phi^{\si\si}_{\be}(0)V_{\al}(z,\bz)\,|\,0\,
\rangle_{z=\frac{i}{2}}^{}
$ w.r.t. $\si$}

Our next aim is to show that 
$
\langle B_{\si}|\,
\Phi^{\si\si}_{\be}(0)V_{\al}(z,\bz)\,|\,0\,
\rangle_{z=\frac{i}{2}}^{}
$
is entire analytic in the variable $\si$.
It is not straightforward to read off the relevant analytic 
properties of $A^{\si}_{\be|\al}$ from \hosoform. We will 
therefore present an alternative integral representation
from which the desired information can be read off easily.

To begin with, let us observe that formula \hosoform\ may be 
written as
\eqn\hosoalt{
\eqalign{
{} & N^{-1} \rho_{\be|\al}^{-1}
A^{\si}_{\be|\al}\;=\;\cr 
& =
e^{-\pi i(2\al+\be-Q)(2\si-Q)}
\frac{S_b(2\al+\be-Q)S_b(\be)}{S_b(2\al)}
F_b\big(2\al+\be-Q,\be,2\al;Q-2\si\big)\;,
}}
where $F_b(A,B,C;x)$ is the b-hypergeometric function 
defined as \qhyp\qPonTe
\eqn\bhypg{
F_b(A,B,C;x)\;\equiv\;\frac{S_b(C)}{S_b(A)S_b(B)}
\int\limits_{\BR+i0}ds\;
\frac{S_b(A+is)S_b(B+is)}{S_b(C+is)S_b(Q+is)}
e^{-2\pi xs}\;  .
}
The following identity was established in \qPonTe:
\eqn\BarnesEuler{
\eqalign{
F_b\big(2\al+\be-Q, & \be,2\al;Q-2\si\big)=\cr
& =\frac{G_b(2\al)}{G_b(\be)G_b(2\al-\be)}
\Psi_b\big(2\al+\be-Q,\be,2\al;2Q-2\si-\be\big),
}}
where $G_b=e^{\frac{\pi i}{2}(x^2-xQ)}S_b(x)$ and 
$\Psi_b(A,B,C;y)$ is defined by the integral
representation
\eqn\Euler{
\Psi_b(A,B,C;y)\;\equiv\;\int\limits_{\BR+i0} ds\;
\frac{G_b(y+is)G_b(C-B+is)}{G_b(y+A+is)G_b(Q+is)}\,
e^{-2\pi sB}\,.
}
From \Euler\ it is straightforward to read off the analytic
properties w.r.t. $\si$ in the same way as explained
in Appendix A.3. We find poles only if 
$Q-\be\pm (Q-2\si)=-nb-mb^{-1}$. These poles are cancelled 
by the multiplication with $g^{\si\si}_{\be}$,
showing that $g^{\si\si}_\be A^{\si}_{\be|\al}$  
is indeed entire analytic in the variable $\si$.

\subsec{Special values of $\langle B_{\si}|\,
\Phi^{\si\si}_{\be}(0)V_{\al}(z,\bz)\,|\,0\,
\rangle_{z=\frac{i}{2}}^{}$}

Of particular importance for us will be certain special values of
\eqn\bulkbddef{
\langle B_{\si}|\,
\Phi_{\si}(0)V_{\al}(z,\bz)\,|\,0\,
\rangle_{z=\frac{i}{2}}^{}\;=\; 
\lim_{\be\ra 2\si} g^{\si\si}_{\be} A^{\si}_{\be|\al}\,.
}
In order to calculate \bulkbddef\
let us begin by considering the 
behavior of the integral in \hosoform\ for $\be\ra 2\si$. 
With the help of eqn. \sbas\ one finds that the integrand
has leading asymptotics $e^{2\pi(2\si-Q\mp(Q-\be))t}$ for 
$t\ra\pm\infty$. It follows that the integral in \hosoform\ 
behaves near $\be=2\si$ as $1/2\pi(2\si-\be)$, which implies
that 
\eqn\hosoasym{
A^{\si}_{\be|\al}\sim \,
\rho_{\be|\al}\,\frac{N}{2\pi(2\si-\be)}+({\rm regular}).
}
Let us furthermore note that \sbdef\ and \sbres\ imply that 
$g^{\si\si}_{\be}$  vanishes at $\be=2\si$ as
\eqn\gpole{
g^{\si\si}_{\be}\sim
\big(\pi \mu\ga(b^2)b^{2-2b^2}\big)^{\frac{\si}{b}}
\frac{\Ga_b(Q-4\si)}{\Ga_b(2Q-4\si)}
\frac{\Ga_b(2\si)\Ga_b(2Q-2\si)}{\Ga_b^2(Q-2\si)}\,
\big[2\pi(2\si-\be)\big]\;.
}
By combining \hosoasym\ and  \gpole\ we are lead to the 
expression
\eqn\hosospec{
\eqalign{
& \langle   B_{\si}|\,
\Phi_{\si}(0)V_{\al}(z,\bz)\,|\,0\,
\rangle_{z=\frac{i}{2}}^{}\;=\;\cr
&N \big(\pi \mu\ga(b^2)b^{2-2b^2}\big)^{\frac{Q-2\al}{2b}}
\frac{\Ga_b(2Q-2\si)\Ga_b(Q-2\si)}{\Ga_b(2Q-4\si)\Ga_b(Q)}
\frac{\Ga(2\al-2\si)\Ga_b(2Q-2\al-2\si)}{\Ga(2\al)\Ga(Q-2\al)}.
}}
Equation \hosopert\ in \S 3.3.
follows from \hosospec\ by means of a straightforward calculation.

\appendix{C}{Analytic continuation of the boundary OPE}

The operator product expansion of the boundary
fields $\Phi^{\sigma_2\sigma_1}_{\alpha_1}(x)$ can be read off from 
the factorization expansion of a four-point function 
\eqn\fourpt{
\big\langle\, \Phi^{\sigma_1\sigma_4}_{\alpha_4}(x_4)\dots
\Phi^{\sigma_2\sigma_1}_{\alpha_1}(x_1)\,\big\rangle.
}
In the case that $\si_k\in\QS$ and $\al_k\in\QS$,  $k=1,\dots,4$  
we know from \PonTe\ and \TLtwo, \S5, that the four-point function \fourpt\ can
be represented by an expansion of 
the following form
\eqn\factor{
\eqalign{
\big\langle\, \Phi^{\sigma_1\sigma_4}_{\alpha_4}(x_4) & \dots
\Phi^{\sigma_2\sigma_1}_{\alpha_1}(x_1)\,\big\rangle=\cr
& =\int_{\QS}d\al\;m_{\al}^{\si_3\si_1}\,
D_{\alpha_{4}\alpha_{3}\alpha}^{\sigma_{4}\sigma_{3}\sigma_{1}}
D_{\alpha\alpha_{2}\alpha_{1}}^{\sigma_{3}\sigma_{2}\sigma_{1}}
\CF_{\al}\big[\,{}^{\al_3}_{\al_4}\;{}^{\al_2}_{\al_1}\,\big](x_4,\dots,x_1).
}}
The measure $m_\al^{\si_2\si_1}$ represents a natural analog of 
the Plancherel measure for the boundary Liouville theory and is given
by
\eqn\Planch{
\eqalign{
m_\al^{\si_2\si_1}\;=\; 
\frac{1}{g_{\al\phantom{Q}}^{\si_2\si_1}g_{Q-\al}^{\si_1\si_2}}
\;=\;{\rm m}(\al)
\frac{1}{2\pi}D(\si_2,\al,\si_1)
\,\,,
}
}
where 
$D(\al_3,\al_2,\al_1)$ is the three-point function 
of the bulk fields $W_{\al}(z,\bar{z}):= \nu_{\al}V_{\al}(z,\bar{z})$,
with $\nu_{\al}\equiv v(P)$ (see \vpdef) for $\al=\frac{Q}{2}+iP$,
and ${\rm m}(\al)=4 \sin \pi b(2\al-Q) \sin\pi b(Q-2\al)$.

In order to disentangle the two main effects it is useful to
focus on the following two particular cases.

\subsec{}
Let us first consider the analytic continuation w.r.t.
the parameters $\si_k$, $k=1,\dots,4$ while keeping the
variables $\al_k$, $k=1,\dots,4$ within the domain
\eqn\restrs{
\eqalign{
{}& |\Re(\al_1-\al_2)|<Q/2, \qquad|\Re(\al_1+\al_2-Q)|<Q/2,\cr
{}& |\Re
(\al_3-\al_4)|<Q/2, \qquad|\Re(\al_3+\al_4-Q)|<Q/2
}}
The results of Appendix A.2.
and \TB, \S7.1, imply that the integrand is meromorphic
in the range under consideration, with poles only coming from the 
measure $m_{\al}^{\si_2\si_1}$.
The discussion of the meromorphic continuation then proceeds along rather
similar lines as in the Appendix A.3, leading to the conclusion that 
\factor\ is replaced by an expression of the following form:
\eqn\factorcont{
\eqalign{
\big\langle\, \Phi^{\sigma_1\sigma_4}_{\alpha_4}(x_4) & \dots
\Phi^{\sigma_2\sigma_1}_{\alpha_1}(x_1)\,\big\rangle=\cr
=& \int_{\QS}d\al\;m_{\al}^{\si_3\si_1}\,
D_{\alpha_{4}\alpha_{3}\alpha}^{\sigma_{4}\sigma_{3}\sigma_{1}}
D_{\alpha\alpha_{2}\alpha_{1}}^{\sigma_{3}\sigma_{2}\sigma_{1}}
\CF_{\al}\big[\,{}^{\al_3}_{\al_4}\;{}^{\al_2}_{\al_1}\,\big](x_4,\dots,x_1)\cr
& \;\;\;
+\sum_{\al\in\QD_{\si_3\si_1}}\frac{1}{|\!| w_{\al;\si_3,\si_1}\!|\! |_{}^2}
D_{\alpha_{4}\alpha_{3}\alpha}^{\sigma_{4}\sigma_{3}\sigma_{1}}
D_{\alpha\alpha_{2}\alpha_{1}}^{\sigma_{3}\sigma_{2}\sigma_{1}}
\CF_{\al}\big[\,{}^{\al_3}_{\al_4}\;{}^{\al_2}_{\al_1}\,\big](x_4,\dots,x_1)\;.
}}
The terms that are summed in the third line of \factorcont\ are produced
by poles of $m_{\al}^{\si_3\si_1}$ that cross the contour of integration
when one continues w.r.t. the variables $\si_1$ and $\si_3$.
The summation is extended over the set 
\eqn\discspec{
\eqalign{
\QD_{\si_3\si_1}=\bigg\{\al\in\QC\; ; \;\al=  Q-
\si_{st}+& nb+mb^{-1}<\frac{Q}{2}  \;,\cr 
 & n,m\in\QZ^{\geq 0}\; ;\; s,t\in\{+,-\}  \,\bigg\},
}
}
where $\si_{st}=s\big(\si_1-\frac{Q}{2}\big)+t\big(\si_3-\frac{Q}{2}\big)$.
The form of the factorization \factorcont\ is easy to understand:
The representations $\CV_{\alpha,c}$ with $\al\in\QD_{\si_3\si_1}$
generate the discrete part in the spectrum 
$\CH_{\sigma_3\sigma_1}^{\rm B}$ \TLbound.
The factorization \factorcont\ is therefore just the result of 
inserting a complete set of intermediate states 
between $\Phi^{\sigma_4\sigma_3}_{\alpha_3}(x_3)$
and $\Phi^{\sigma_3\sigma_2}_{\alpha_2}(x_2)$, taking into 
account that $\Phi^{\sigma_3\sigma_2}_{\alpha_2}(x_2)
\Phi^{\sigma_2\sigma_1}_{\alpha_1}(x_1)$ creates states 
within $\CH_{\sigma_3\sigma_1}^{\rm B}$ when acting on the vacuum.

\subsec{}
Let us now consider the analytic continuation w.r.t.
the parameters $\al_k$, $k=1,\dots,4$ while keeping the
variables $\si_k$, $k=1,\dots,4$ within the domain
\eqn\restrs{
|\Re(\si_1-\si_3)|<Q/2, \qquad|\Re(\si_1+\si_3-Q)|<Q/2\;.
}
We now only have to consider the poles of the three-point functions
$D_{\alpha_{4}\alpha_{3}\alpha}^{\sigma_{4}\sigma_{3}\sigma_{1}}$ and 
$D_{\alpha\alpha_{2}\alpha_{1}}^{\sigma_{3}\sigma_{2}\sigma_{1}}$.
The discussion is largely parallel to \TB, \S7.1,
allowing us to conclude that the for generic values of 
$\al_k$, $k=1,\dots,4$ we get a factorization of the following form:
\eqn\factorconttwo{
\eqalign{
\big\langle\, \Phi^{\sigma_1\sigma_4}_{\alpha_4}(x_4) & \dots
\Phi^{\sigma_2\sigma_1}_{\alpha_1}(x_1)\,\big\rangle=\cr
=& \int_{\QS}d\al\; m_{\al}^{\si_3\si_1}\,
D_{\alpha_{4}\alpha_{3}\alpha}^{\sigma_{4}\sigma_{3}\sigma_{1}}
D_{\alpha\alpha_{2}\alpha_{1}}^{\sigma_{3}\sigma_{2}\sigma_{1}}
\CF_{\al}\big[\,{}^{\al_3}_{\al_4}\;{}^{\al_2}_{\al_1}\,\big](x_4,\dots,x_1)\cr
& \;\;\;
+\sum_{\al\in\QD_{\al_2\al_1}} m_{\al}^{\si_3\si_1}
D_{\alpha_{4}\alpha_{3}\alpha}^{\sigma_{4}\sigma_{3}\sigma_{1}}
d_{\alpha\alpha_{2}\alpha_{1}}^{\sigma_{3}\sigma_{2}\sigma_{1}}
\CF_{\al}\big[\,{}^{\al_3}_{\al_4}\;{}^{\al_2}_{\al_1}\,\big](x_4,\dots,x_1)\cr
& \;\;\;
+\sum_{\al\in\QD_{\al_3\al_4}} m_{\al}^{\si_3\si_1}
d_{\alpha_{4}\alpha_{3}\alpha}^{\sigma_{4}\sigma_{3}\sigma_{1}}
D_{\alpha\alpha_{2}\alpha_{1}}^{\sigma_{3}\sigma_{2}\sigma_{1}}
\CF_{\al}\big[\,{}^{\al_3}_{\al_4}\;{}^{\al_2}_{\al_1}\,\big](x_4,\dots,x_1)\;,
}}
where $d_{\alpha_{3}\alpha_{2}\alpha_1}^{\sigma_{3}\sigma_{2}\sigma_{1}}$
is the relevant residue of 
$D_{\alpha_{3}\alpha_{2}\alpha_1}^{\sigma_{3}\sigma_{2}\sigma_{1}}$,
\eqn\smalld{
d_{\alpha_{3}\alpha_{2}\alpha_1}^{\sigma_{3}\sigma_{2}\sigma_{1}}\;=\;
2\pi i \mathop{\rm Res}_{\al_3\in\QD_{\al_2\al_1}}\!\!
D_{\alpha_{3}\alpha_{2}\alpha_1}^{\sigma_{3}\sigma_{2}\sigma_{1}}\;.
}
In the general case one clearly finds discrete terms of 
the two different types appearing in \factorcont\ and \factorconttwo\ 
respectively.

\appendix{D}{The limit $\si\ra 0$}

In this appendix we present the proofs of the statements presented in 
\S2.6.

\subsec{Proof of \zerolimone}

To begin with, let us note that the coefficient $G_{\al}^{\si_2\si_1}$
is indeed finite in the limit $\si_i\ra 0$, $i=1,2$ 
for generic values of 
$\al$, unlike $g_{\al}^{\si_2\si_1}$. It follows that expectation values
involving the boundary fields $\tilde{\Phi}^{\sigma_2\sigma_1}_{\alpha}(x)$
will be generically finite. The boundary fields 
$\Phi_{\si}(x)$ 
which create the bound state with lowest conformal dimension 
do not need to be renormalized before taking $\si\ra 0$.
This can be seen from our expression \twoptphi\ for the two-point 
function of the fields $\Phi_{\si}(x)$. 
Let us therefore introduce the notation 
\eqn\limitfields{
\tilde{\Phi}_{\al}(x)\;:=\;
\lim_{\si\ra 0}\tilde{\Phi}^{\sigma\sigma}_{\alpha}(x),\qquad
\Phi_0(x)\;:=\;\lim_{\si\ra 0}\Phi_{\si}(x)
}
for the relevant boundary fields in the $\si=0$ boundary Liouville theory
as well as 
\eqn\dtildedef{
\eqalign{
D_{\ta_3\ta_2\ta_1}\,:=\, & \lim_{\si\ra 0}\,\big\langle\,
\tilde{\Phi}^{\sigma\sigma}_{\alpha_3}(\infty)
\tilde{\Phi}^{\sigma\sigma}_{\alpha_2}(1)
\tilde{\Phi}^{\sigma\sigma}_{\alpha_1}(0)\,\big\rangle
\cr 
D_{\ta_3\ta_2|0}\,:=\, & \lim_{\si\ra 0}\,  \langle\,
\tilde{\Phi}^{\sigma\sigma}_{\alpha_3}(\infty)
\tilde{\Phi}^{\sigma\sigma}_{\alpha_2}(1)
\Phi_{2\si}^{\sigma\sigma}(0)\,\big\rangle
\cr
D_{\ta_3|0,0}\,:=\, & \lim_{\si\ra 0}\,\big\langle\,
\tilde{\Phi}^{\sigma\sigma}_{\alpha_3}(\infty)
\Phi_{2\si}^{\sigma\sigma}(1)
\Phi_{\si}(0)\,\big\rangle\cr 
D_{0,0,0}\,:=\, &   \lim_{\si\ra 0}\,\big\langle\,
\Phi_{2\si}^{\sigma\sigma}(\infty)
\Phi_{2\si}^{\sigma\sigma}(1)
\Phi_{2\si}^{\sigma\sigma}(0)\,\big\rangle
}
}
for the corresponding three point functions.
By using \spOPEcoeffone\ and \spOPEcoefftwo\ one may verify that
\eqn\limitvan{
D_{\ta_3\ta_2|0}\;=\;0,
\quad{\rm and}\quad
D_{\ta_3|0,0}\;=\;0,
}
whereas $D_{\ta_3\ta_2\ta_1}$ and $D_{0,0,0}$
will be finite. We furthermore
need to discuss the structure of the OPE for $\si\ra 0$.
This may be read off from \factorcont\ if one takes into account the
renormalization relating the boundary fields
$\tilde{\Phi}^{\sigma_2\sigma_1}_{\alpha}(x)$ and
$\Phi^{\sigma_2\sigma_1}_{\alpha}(x)$, equations 
\bdrenormtwo\ and \bigGsmallg, as well as equation \limitvan\ expressing
the vanishing of mixed three point functions.
We thereby infer that the operator product expansion
of $\tilde{\Phi}_{\al_2}(x_2)\tilde{\Phi}_{\al_1}(x_1)$
does {\it not} contain $\Phi_0(x_1)$.
By taking into account that the conformal block $\CF_{\al}$ has 
a first order pole at $\al=0$ one may furthermore 
observe that four-point functions which contain
$\tilde{\Phi}_{\al_2}(x_2)\Phi_0(x_1)$ vanish. 
This, together with \limitvan\ 
is good enough to infer the vanishing of all mixed 
correlation functions of only boundary fields. 

To conclude, let us show that $\Phi_0(x_1)$ indeed 
represents the projector onto the sector $\CV_{0,c}$. 
Let us first note that $\Phi_0(x)$ is a primary field with
conformal dimension $0$. Vanishing of all mixed correlation
functions implies that the fusion rules for decoupling of
the null vector in $\CV_{0,c}$ are satisfied, so that 
$\pa_x\Phi_0(x)=0$.  Let us finally observe that by comparing 
\twoptphi\ and \spOPEcoeffthree\ in the limit $\si\ra 0$ 
one finds that
\eqn\threepttwopt{
\langle\,
\Phi_0
\Phi_0
\Phi_0\,\rangle
\;=\;\langle\,
\Phi_0
\Phi_0\,\rangle\, .
}
We conclude that $\Phi_0^2=\Phi_0$, which means that 
$\Phi_0$ has the correct normalization for representing the
projection onto $\CV_{0,c}$. 

\subsec{Proof of \projprop}

By using the 
operator product expansion
of the bulk fields $V_{\al}$ one may reduce the proof of 
formula \projprop\ to the following
result:
\eqn\projlem{
\lim_{\si\ra 0}\,\lim_{\be\ra 2\si}\;\langle B_{\si}|\,
\Psi^{\si\si}_{\be}(0)V_{\al}(z,\bz)\,|\,0\,\rangle_{z=\frac{i}{2}}^{}
\;=\; \langle B_{\rm D}|\,
V_{\al}(z,\bz)\,|\,0\,
\rangle_{z=\frac{i}{2}}^{}
}
if $
\al=\frac{Q}{2}+iP$. Equation \projlem\ follows straightforwardly
from \hosospec, taking into account that 
$N=-2^{\frac{5}{4}}\sqrt{\pi}$ and that
$\langle B_{\rm D}|\,
V_{\al}(z,\bz)\,|\,0\,
\rangle_{z=\frac{i}{2}}^{}=\sqrt{2\pi}\Psi_{\rm ZZ}(P)$.

\noindent

\listrefs
\end

\bye
\end